\documentclass{article}
\usepackage{amsmath}
\usepackage{amssymb}
\usepackage[usenames]{color}
\usepackage{booktabs}
\usepackage{multirow}
\usepackage{comment}
\usepackage{paralist}
\usepackage{subcaption}
\usepackage[pagewise]{lineno}%\linenumbers %% Mette i numeri alla riga
\usepackage[ruled,vlined]{algorithm2e}
\usepackage{graphicx}  
\usepackage{epstopdf} %This is to transfer .eps figure to .pdf figure; please compile your paper using PDFLeTex or PDFTeXify.
\usepackage[colorlinks=true]{hyperref}
\hypersetup{urlcolor=blue, citecolor=red}
\graphicspath{{figures/}}

\newcommand{\R}{\mathbb{R}}
\DeclareMathOperator{\prox}{{prox}}
\DeclareMathOperator*{\argmin}{argmin}
\newcommand\silviacorr[1]{\textcolor{black}{#1}}

\title{Explainable bilevel optimization: an application to the Helsinki deblur challenge}

\begin{document}
\maketitle

% Enter the first author's name and address:
%\Huge{}
\centerline{\scshape Silvia Bonettini, Giorgia Franchini, Danilo Pezzi and Marco Prato}
\medskip
\centerline{silvia.bonettini@unimore.it,giorgia.franchini@unimore.it}
\centerline{danilo.pezzi@unimore.it,marco.prato@unimore.it}
{\footnotesize
% please put the address of the first author
 \centerline{Dipartimento di Scienze Fisiche, Informatiche e Matematiche}
 \centerline{Universit\`a di Modena e Reggio Emilia }
 \centerline{Via Campi 213/b - 41125 Modena (MO), Italy}
 \centerline{All the authors are members of the INdAM research group GNCS}
} % Do not forget to end the {\footnotesize by the sign }

\bigskip

% The name of the associate editor will be entered by an editorial staff
% "Communicated by the associate editor name" is not needed for special issue.
%\centerline{(Communicated by the associate editor name)}

%The abstract of your paper
\begin{abstract}
In this paper we present a bilevel optimization scheme for the solution of a general image deblurring problem, in which a parametric variational-like approach is encapsulated within a machine learning scheme to provide a high quality reconstructed image with automatically learned parameters. The ingredients of the variational lower level and the machine learning upper one are specifically chosen for the Helsinki Deblur Challenge 2021, in which sequences of letters are asked to be recovered from out-of-focus photographs with increasing levels of blur. Our proposed procedure for the reconstructed image consists in a fixed number of FISTA iterations applied to the minimization of an edge preserving and binarization enforcing regularized least-squares functional. The parameters defining the variational model and the optimization steps, which, unlike most deep learning approaches, all have a precise and interpretable meaning, are learned via either a similarity index or a support vector machine strategy. Numerical experiments on the test images provided by the challenge authors show significant gains with respect to a standard variational approach and performances comparable with those of some of the proposed deep learning based algorithms which require the optimization of millions of parameters.
\end{abstract}
Keywords: {Bilevel optimization, Blind deconvolution methods, Helsinki deblur challenge, Performance predictor, GreenAI}

\section{Introduction}
\label{sec:1}
In a general image deblurring problem, we assume that the data $f\in\R^n$ is a blurred noisy version of some true image $g$
\begin{equation*}
H g + \eta = f,
\end{equation*} 
where $H\in\R^{n\times n}$ represents the blurring operator and $\eta$ denotes the statistical, unknown noise affecting the data. In general, $H$ is a structured matrix defined in such a way that the product $Hu$ corresponds to a convolution between the image $u$ and a given kernel $h$ representing the Point Spread Function (PSF) of the imaging system employed to measure the data. The deblurring (or deconvolution) problem consists in finding an approximation of $g$, given the blurred image $f$ and, possibly, some information on the system PSF. If the blurring kernel $h$, underlying the matrix $H$, is completely unknown and it has to be inferred together with $g$, the resulting problem is a blind deconvolution one \cite{Levin2011}.

Since the PSF $h$ usually represents a low--pass filter, the matrix $H$ is, at best, very ill conditioned and directly solving the inverse problem $Hu=f$, even when it is feasible, leads to unmeaningful solutions. On the other side, the variational approach consists in designing and solving an optimization problem whose solutions are a good approximation of the unknown image $g$. \silviacorr{In general, a variational model is the set composed by the objective function, i.e., the function to be minimized, and the possible constraints.} In the variational models arising in image restoration applications, the objective function, called also energy functional, encompasses different kinds of information: the nature of the noise introduced in the acquisition process, geometrical and/or analytical properties on the image content and physical constraints on the pixel values. Usually, in all image reconstruction problems, and more generally inverse problems, the energy functional, besides the data, depends on a set of parameters; they may simply reduce to tuning parameters balancing the relative weights of the different terms in the functional but can also represent more complicate structures of the functionals themselves. In the following, we will denote by $\theta\in\R^p$ the vector collecting all parameters of the energy functional, and explicit its dependence on $\theta$ and on the data $f$ by writing it as $E_f(u,\theta)$.

Within these settings, the stages of the classical variational approach are the following ones.
\begin{itemize}
    \item Modeling: define the fit-to-data and the regularization terms, according to the noise statistics and the prior information.
    \item Parameters selection: choose a set of parameters $\theta\in\R^p$.
    \item Optimization: compute a solution of the following minimization problem
    \begin{equation*}
\tilde u(\theta) \in \argmin_{u\in\R^n} E_f(u,\theta).
\end{equation*}
\end{itemize}

One of the main difficulties of the above procedure lies in the second stage: indeed, even if discrepancy principles with well established theoretical motivation are available for selecting the regularization parameter in some instances of energy functionals, explicit guidelines in the general case are an open problem. The need of manually tuning the underlying set of parameters imposes that their number must be small and the dependency on the functional structure simple. Once the minimization problem has been completely settled, a solution is computed by a suitable optimization method.

In order to overcome these difficulties, in the last decade a new research field, based on deep learning techniques, enormously grew up. However, the lack of interpretability has become a main barrier of deep learning in its wide acceptance in mission-critical applications \cite{Fan2021OnIO}. In addition to this lack of interpretability, deep learning models are heavily influenced by architectural choices, the design of which is particularly computationally expensive. The extensive repeated testing required to design a good architecture, each of which requires training millions of parameters, is in pronounced contrast to the recently developing strand of GreenAI \cite{Schwartz2020}. 

Besides deep learning techniques, another recent strategy to overcome the difficulties of the classical approach by mixing variational and machine learning techniques consists in the bilevel optimization approach \cite{Arridge-etal-2019,Chen2014,Franceschi2018,Kunisch-Pock-2013}. In this framework, we have to assume that a dataset of samples $\{g^{(s)}, f^{(s)}\}_{s=1}^S$ is available, where $f^{(s)}$ is a noisy blurred version of $g^{(s)}$. Then, a loss function $\ell(\cdot,\cdot)$ is adopted to measure the distance, or the similarity, between two images and the set of parameters $\theta\in\R^p$ is computed by solving the following bilevel optimization problem:
\begin{equation*}
    \begin{array}{ll}\displaystyle \min_{\theta\in\R^p} & \displaystyle\sum_{s=1}^S\ell(\tilde u^{(s)}(\theta),g^{(s)})\\
    \mbox{s.t. } & \tilde u^{(s)}(\theta) = \displaystyle\argmin_{u\in\R^n}E_{f^{(s)}}(u,\theta).
    \end{array}
\end{equation*}

However, solving the above problem can be clearly impractical for both theoretical and computational reasons, mainly due to the fact that the minimization problem providing $\tilde u^{(s)}(\theta)$, in general, can not be solved in closed form and its approximate solution through iterative procedures is a computational demanding task.

A natural development of the bilevel approach is the unrolling technique \cite{Arridge-etal-2019,Hershey2014,Monga2021,Ochs-etal-2015}, where the lower level problem, i.e., the minimization of the energy functional, is replaced by a fixed, finite number $K\in\mathbb{N}$ of iterations of an optimization method applied to it. In practice, the optimization method, stopped after $K\in\mathbb{N}$ iterations, and the underlying variational model are considered as an image restoration procedure whose parameters have to be optimized with respect to a given dataset of images. The realization of these ideas is summarized below.   
\begin{itemize}
%\item Data: $\{g^{(s)}, f^{(s)}\}_{s=1}^S$, where $f^{(s)}$ is a blurred noisy %version of $g^{(s)}$, all with the same blurring kernel and the same noise %statistics.
\item Modeling: define an energy functional $E_f(u,\theta)$ based on noise statistics and prior information.
\item Unrolling: choose an iterative optimization method to be applied to problem $\min_{u\in\R^n}E_f(u,\theta)$. We will denote by $\mathcal A(f,\theta)$ all the operations needed to compute the new iterate from the previous one(s) for a given parameter array $\theta$. 
%        \begin{equation*}
%            u^{(k+1)} = \mathcal{A}(u^{(k)},\theta)
%        \end{equation*}
\item Parameters learning: solve the following minimization problem 
    \begin{equation}\label{prob:unrolling}
        \begin{array}{ll}\displaystyle\min_{\theta\in\R^p} &\displaystyle\sum_{s=1}^S\ell( u^{*,s}(\theta), g^{(s)})\\
        \mbox{s.t.} & u^{*,s}(\theta) = {\mathcal A}^K(f^{(s)},\theta),
        \end{array}
    \end{equation}
    where $K$ is a prefixed number of iterations, and ${\mathcal A}^K =\overbrace{{\mathcal A}\circ {\mathcal A}\circ \cdots {\mathcal A}}^{K}$. For simplicity, we also assume that the starting point of the iterations is the data itself. As concerns the function $\ell(\cdot,\cdot)$, in a supervised setting it typically represents a loss function measuring the distance, or the similarity, between two images. However, as we will see in Section \ref{sec_loss}, it might also be chosen as a figure of merit of the restored image one wants to obtain, independently of a set of ground truth images.
\end{itemize}
%The learning procedure described above aims to compute the set of parameters $\theta^*\in\R^p$ such that the image obtained after $K$ steps of the method $\mathcal A$ applied to $\min_{u\in\R^n}E(u,\theta)$ with the data itself as initial guess minimizes the objective function $\ell$ over a set of training images.\\

Once a solution $\theta^*$ of \eqref{prob:unrolling} has been computed, the restoration of a blurred noisy image $f$ can be obtained by applying $K$ times the operations in $\mathcal{A}$ to $f$ with the learned parameters $\theta^*$:
\begin{equation*}
    \tilde u = {\mathcal A}^K(f,\theta^*).
\end{equation*}

Further developments of the unrolling ideas aim to increase the model capacity, i.e., the number of parameters and, in turn, the capability to capture more complicated features in the underlying model. For example, several authors allow the set of parameters $\theta$ to change at each unrolled iteration \cite{Bertocchi-etal-2019,Chen-Pock-2017}, up to replace entire parts of the model and/or of the restoration procedure with a Deep Neural Network (DNN) \cite{Kobler-etal-2020,Kobler-etal-2021}.

We are perfectly aware that well designed deep learning techniques lead to the most outstanding results, far better than traditional approaches. \silviacorr{However, besides the need of a very large amount of training data, the price to pay is a poorly explainable model, where for {\em{explainability}} we mean existence of theoretical guidelines for hyperparameters tuning, availability of stable numerical tools for the computation of the solution, and possibility to extract useful information from the algorithm's output. An example of what we consider an explainable algorithm is the Support Vector machines for Regression methodology \cite{Cristianini2000,Drucker-etal-1996}, in which a) the hyperparameters have a precise mathematical meaning and a direct correspondence with the fitting capability of the model, b) its training consists in solving a constrained QP problem by means of a deterministic optimization method, usually a projected gradient or interior point algorithm, which have well-established theoretical convergence properties and can be implemented with adaptive strategies to automatically select its own parameters, and c) its output are the couples of Lagrange multipliers associated to each training example and they provide information on the model accuracy, number of outliers, etc. All these considerations drop in the case of a neural network. First of all, there are no clear and theoretically justified guidelines to follow for setting most part of their many hyperparameters (e.g., how many layers? Which size of the convolutional filters? How choosing the activation functions?). As for the numerical viewpoint, the optimization method applied in the training phase is stochastic and its hyperparameters, such as the learning rate and the minibatch size, have to be chosen almost empirically, although some practical rules have been proposed in the recent years \cite{HYPER1,HYPER2}. Finally, from the millions of parameters to be optimized, it is quite hard to extract meaningful information.} For all these reasons, we believe that non--deep unrolling techniques deserve further analysis, especially in cases, like image deblurring, where there exists a well established literature on the problem itself and on the mathematical tools to handle it.

Based on these motivations, in the present paper we propose an unrolling algorithm especially tailored for image deblurring. Although the main ideas behind the unrolling approach can be applied to any image deblurring problem as well as to a variety of other imaging problems, we will focus our attention (and, consequently, our model and numerical tests) on the recently proposed Helsinki Deblur Challenge (HDC)\footnote{Link to the challenge website: \url{https://www.fips.fi/HDC2021.php}}. As the name suggests, the goal of the challenge was to design a deconvolution algorithm able to perform well on a very specific set of text images (more details on HDC are given in Section \ref{sec:HDC}). Even if the blurring kernel is unknown, the challenge competitors were provided by a certain number of samples of the images to be restored, therefore learning techniques are a natural choice to exploit the availability of this data.

The paper is organized as follows. After a brief description of the HDC rules in Section \ref{sec:HDC}, in Sections \ref{sec2} and \ref{sec3} we introduce, respectively, the energy functional to be minimized, and the unrolling scheme adopted to perform the minimization in the lower level. Some specific choices we made are justified in the HDC framework. In Section \ref{sec_loss} we then describe our choices for the objective function of the upper level aimed at optimizing the set of parameters, while in Section \ref{sec_tests} we show our results on the HDC images.

\subsubsection*{Notations and basic definitions} In most part of the paper we consider a bidimensional image of size $N\times M$ pixels as a vector in $\R^n$, where $n= NM$, assuming that the vectorization proceeds columnwise. However, when it is more convenient, we also employ a two index notation. We indicate with $u_i$ or $[u]_i$ the $i$-th component of the vector $u$. Similarly, $W_{ij}$ or $[W]_{ij}$ denote the entry on the $i$-th row, $j$-th column of the matrix $W$. The notation $u\leq u_{max}$, where $u,u_{max}\in\R^n$ indicates that all components of $u$ are smaller or equal to the corresponding component of $u_{max}$. We indicate the non--negative (positive) orthant of the $n$-dimensional space as $\R^n_{\geq 0}$ ($\R^n_{>0}$, respectively). The 2D convolution is denoted by the asterisk '$*$' and it is assumed with reflective boundary conditions. Given a closed, convex set $\Omega\subseteq\R^n$, the indicator function associated to it is defined as
\begin{equation*}
    \iota_{\Omega}(u) = \left\{\begin{array}{ll}
    0&\mbox{ if } u\in\Omega \\
    +\infty &\mbox{ otherwise.}
    \end{array}\right.
\end{equation*}

\section{The Helsinki Deblur Challenge rules}\label{sec:HDC}

The purpose of the challenge was to deconvolve a set of text images of size $1460\times 2360$ pixels with black characters over a light background, written in two different fonts, Verdana and Times New Roman. Each image represents a random string of text on three different lines. The images were partitioned into 20 levels (also referred to as steps), one for each intensity of blur. The higher the level, the more the images were degraded. Also, for each level, a set of 200 images for each font, partitioned in two subset, was provided. All images were obtained by two digital cameras, denoted by CAM1 and CAM2, which shot the same e-ink screen displaying the string of text. The first camera was on-focus and its images, collected in the first subset of data, are a sort of ground truth. The second subset contains the images from CAM2, which was misfocused and, therefore, they are blurred, noisy and, in addition, they suffer also of other optical distortions. In practice, the dataset contains a good quality image and a blurred, noisy, warped version of it, for each string of text and for each blur level.

A different test set, containing 40 images per level and built with the same modalities, was released after the challenge ended. In the HDC rules, the quality of an image is measured by passing it to an Optical Character Recognition software (OCR) provided together with the initial dataset, in terms of the percentage of correctly recognized characters of the central text line.

The algorithms presented by competitors were applied to the test set and an average score of at least 70 out of 100 of correctly recognized characters by the OCR software meant that the level was successfully passed. The CAM2 images of the first 3 levels actually pass the OCR test without need of any processing, while in the last levels the amount of blur is really extreme. The OCR score is more sensitive to blur rather than noise or warp, therefore it can be reasonably adopted as a quality measure for images of this kind.

\section{Modeling the energy functional}\label{sec2}

As mentioned in the previous section, the Bayesian approach for the solution of an inverse problem is to assume a statistics on the noise affecting the data and a prior on the unknown, and to maximize the posterior probability provided by the Bayes formula \cite{Arridge-etal-2019,Bertero2008}. After some standard mathematical transformations, the resulting energy functional to be minimized is given by the sum of a data fidelity term and a regularization functional
\begin{equation*}
E_f(u) = \mathcal{D}(Hu,f) + \mathcal{R}(u).
\end{equation*}

As for the fit-to-data term, one of the more common choices is the least squares functional
\begin{equation*}
\mathcal{D}(Hu,f)=\frac 1 2 \|H u - f\|^2,
\end{equation*}
which corresponds to the assumption of a Gaussian distribution on the noise on the data, while other distance or distance-like functions can be defined in the Bayesian framework according to different noise statistics, as the Poisson, Cauchy, Laplace or salt-and-pepper ones (see e.g. \cite{Bertero2018,Kosko2006}).

On the other side, the regularization functional $\mathcal{R}$ is selected taking into account any prior information on the true image, enforcing some desired features on the minimizers of $E_f$. For example, the Tikhonov regularization $\mathcal{R}(u)= \frac 1 2\|u\|^2$ promotes smooth solutions, while the sharpness of the edges can be preserved by the Total Variation (TV) functional
\begin{equation}\label{TV0}
\mathcal{R}(u)=    \sum_{i=1}^n \|\nabla_iu\|,
\end{equation}
where $\nabla_i\in\R^2$ represents the discrete gradient of the image $u$
at pixel $i$, and $\|\cdot\|$ denotes the $\ell_1$ or the $\ell_2$ norm (corresponding to the isotropic and anisotropic version of the functional respectively).  Moreover, in imaging problems, only non-negative solutions have physical meaning, and this constraint can be formally imposed by including in the regularization term the indicator function of the non-negative orthant $\iota_{\R^n_{\geq 0}}(u)$.

Both the fit-to-data term and the regularizer might depend on a set of parameters. In blind (or myopic) deconvolution problems, for example, the blurring model is not explicitly available and a parametric form of blurring kernel is in general assumed \cite{Christou1999,Conan1998}. As for the regularization part of the energy functional, we may have the (simplest) case where the parameters are weights balancing its relevance w.r.t. the data fidelity term, but more complex priors, as in the case e.g. of higher order filter-based Markov Random Field (MRF) models \cite{Chen2014,Roth2009}, require the estimation of a very large number of parameters.

According to the notation introduced in the previous section, we will therefore consider the minimization of an energy function that writes as
\begin{equation*}
E_f(u,\theta) = \mathcal{D}(H(\theta)u,f,\theta) + \mathcal{R}(u,\theta).
\end{equation*}
where $\theta\in\R^p$ denotes the set of all parameters defining the blurring kernel and the regularization term. Even if the most part of the analysis in Section \ref{sec3} can be applied in these general settings, in the following section we describe and motivate the specific choices of the model we made in view of the application to the Helsinki deblur challenge.

\subsection{The HDC model}

It is well understood that a general purpose prior/regularizer does not exist: on the contrary, it must be defined using as much information one has on the data and on the target solution of his problem. In this perspective, the most relevant aspects about our reference problem are the following ones 
\begin{itemize}
    \item the data suffer from out-of-focus blur.
    \item the ideal target of the restoration process is a binary, piecewise constant image.
\end{itemize}

The out-of focus blur kernel can be modeled as the characteristic function of a disc \cite{Bertero1998}: increasing the radius $r$ of the disc produces images which are more and more blurred. We consider the same discretization of the characteristic function of the disc implemented in the Matlab function \verb"fspecial" and, here and in the following, we will denote by $H(r)$ the matrix representing the convolution with this kernel.

As for the regularization terms, the TV functional \eqref{TV0} is well suited when seeking for piecewise constant solutions. More precisely, in order to avoid nonsmoothness, we adopt the following approximation of the functional in \eqref{TV0}
\begin{equation}\label{TV}
TV(u;\delta) = \sum_{i=1}^n \sum_{j=1}^2\sqrt{\left[\nabla_i u \right]_j^2+\delta^2},
\end{equation}
where $\delta$ is a positive scalar. The above functional, for small values of $\delta$, can be considered as an approximation of the standard TV, but it has been analyzed also in the more general context of MRF priors \cite{Zanella-etal-2009}, whose general form is
\begin{equation}\label{MRF}
    \sum_{i=1}^n\sum_{j=1}^{J} \phi(\left[\kappa_j*u \right]_i,\delta),
\end{equation}
where $\phi(\cdot,\delta):\R\to \R_{\geq 0}$ is a weighting function depending on the parameter $\delta$, while $\kappa_j$, $j=1,..,J$ are convolution kernels with zero mean. The TV function \eqref{TV} corresponds to the settings $\phi(t,\delta) = \sqrt{t^2+\delta^2}$, $J=2$, with $\kappa_1$ and $\kappa_2$ representing the finite difference operators in the horizontal and vertical directions. Another interesting setting of the MRF prior is proposed in \cite{Chen2014,Chen-Pock-2017} in the framework of natural images restoration: in this case, a larger number of kernels (from 25 to 80) is adopted and their components are learned by means of a bilevel optimization strategy. These settings, with a proper choice of the function $\phi$, are motivated by statistical arguments and showed to be very well suited for capturing the complicated dynamics of natural images. Even if all the subsequent analysis in the present paper can be applied also when the energy functional includes the term in \eqref{MRF}, we believe that the HDC data and the features of the target images do not require a prior with such a complex form. Then, we focus on the simplest case \eqref{TV} for promoting edge sharpness and, on the other side, we include an additional term to model the other desired features of the output. 

In particular, as a prior for binary images, assuming that the data are scaled in a proper way, we propose to adopt the combination of a box constraint over the $n$-rectangle $[0,1]^n$ 
%\begin{equation}\label{box}
%    \iota_{[0,1]^n}(u) = \left\{\begin{array}{ll}
%    0&\mbox{ if } u\in [0,1]^n\\
%    +\infty &\mbox{ otherwise.}
%    \end{array}\right.
%\end{equation}
and the following bimodal function
\begin{equation}\label{bimodal}
B(u) = \frac 1 2 \sum_{i=1}^nu_i(1-u_i).
\end{equation}
The above functional is concave in $[0,1]^n$, since $\nabla^2B(u) = -I$ and it forces the pixels to achieve the bounds of the feasible region. Finally, the regularization term is defined as a combination of \eqref{TV}--\eqref{bimodal}, with the constraints in the $n$-rectangle.

In summary, the energy functional is defined as
\begin{equation}\label{energy}
E_f(u,\theta) = \frac 1 2 \|H(r)u - f\|^2 + \rho B(u) +  \gamma TV(u;\delta),
\end{equation}
where 
\begin{equation}\label{thetadef}
\theta = (r,\rho,\gamma,\delta)^T
\end{equation}
is the corresponding set of parameters and we consider the following constrained variational model
\begin{equation}\label{minE}
    \min_{0\leq u\leq 1} \ E_f(u,\theta).
\end{equation}

\section{Unrolling technique}\label{sec3}

The Fast Iterative Soft Thresholding Algorithm (FISTA) \cite{Beck-Teboulle-2009b,Chambolle-Dossal-2015} is one of the most popular and effective optimization methods which can be applied to an optimization problem of the form
\begin{equation}\label{prob:composite}
    \min_{u\in\R^n} \psi_0(u)+\psi_1(u),
\end{equation}
where $\psi_0:\R^n\to \R$ is a convex functions which is continuously differentiable on a convex set $\Omega \subseteq \R^n$ and $\psi_1:\R^n\to \R\cup\{+\infty\}$ is a lower semicontinuous convex function with $\mbox{dom}(\psi_1)\supseteq \Omega$ . In the recent literature, several variants of FISTA have been proposed. In this paper we consider the following iteration:
\begin{equation*}
\begin{array}{ll}
%\mbox{for }  k = 0,1,...\\
 %\left\lfloor
 \begin{array}{lcl}
						  v^{(k)} &=&P_{\Omega}( u^{(k)} + \beta_k(u^{(k)}-u^{(k-1)}))\\[0.2cm]
							 u^{(k+1)}     &=& \prox_{\alpha_k\psi_1}(v^{(k)}-\alpha \nabla \psi_0(v^{(k)})),
					    \end{array}%\right.\\
\end{array}
\end{equation*}
where $u^{(-1)}=u^{(0)}\in\Omega$ are the starting points, $P_\Omega(\cdot)$ denotes the orthogonal projection onto the set $\Omega$ and $\prox_{\psi}$ is the proximal operator defined as
\begin{equation*}
    \prox_{\psi}(z) = \arg\min_{u\in\R^n} \psi(u) + \frac 1 2 \|u-z\|^2.
\end{equation*}
Moreover, $\alpha_k$ and $\beta_k$ are the steplength and extrapolation parameters, respectively. The convergence properties of the above iteration have been established in \cite{Bonettini-Rebegoldi-Ruggiero-2018}, with suitable choices of $\alpha_k$ and $\beta_k$. More general results for FISTA-like methods in nonconvex settings can be found in \cite{Ochs-Pock-2019}.

In the framework of the unrolling techniques outlined in Section \ref{sec:1}, we consider the FISTA iteration applied to the constrained minimization of the energy functional \eqref{energy}. Indeed, problem \eqref{minE} can be cast in the form \eqref{prob:composite} by setting $\Omega = [0,1]^n$, $\psi_0 = E_f$ and $\psi_1 = \iota_{[0,1]^n}$. As a consequence of this, $\prox_{\psi_1}$ reduces to the orthogonal projection onto the $n$-rectangle $[0,1]^n$. This kind of constraint can be easily handled by optimization methods, since the projection operator is available in closed form as $P_{[0,1]^n}(u)= \max\{\min\{u,1\},0\}$. However, problem \eqref{prob:unrolling} is much easier to handle if the algorithm rule $\mathcal A$ is smooth. For such reason, in this framework it is quite usual to replace the Euclidean projector and/or the proximity operator with a smooth projection-like function, possibly corresponding to a given metric \cite{Auslender-Teboulle-2009,Bertocchi-etal-2019,Gregor-LeCun-2010}. Here we propose the following projection-like function
\begin{equation}\label{proj}
    \Pi(u) = \left\{\begin{array}{ll}
    %0 & \mbox{if } u\leq0\\
    %1 & \mbox{if } u \geq 1\\
    %u & \mbox{if } u\in [\epsilon,1-\epsilon]\\
    \max\{\min\{u,1\},0\}& \mbox{if } u\in [-\infty,0]\cup[\epsilon,1-\epsilon]\cup [1,+\infty]\\
    \left(2-\frac {u} \epsilon\right)\frac {u^2} \epsilon&\mbox{if } u\in [0,\epsilon]\\
    1 - \left(2-\frac {1-u} \epsilon\right)\frac {(1-u)^2} \epsilon&\mbox{if } u\in [1-\epsilon,1]
    \end{array}
    \right.
\end{equation}
where $\epsilon$ is a positive parameter. 

\begin{figure}
    \centering
    \includegraphics[scale=0.5]{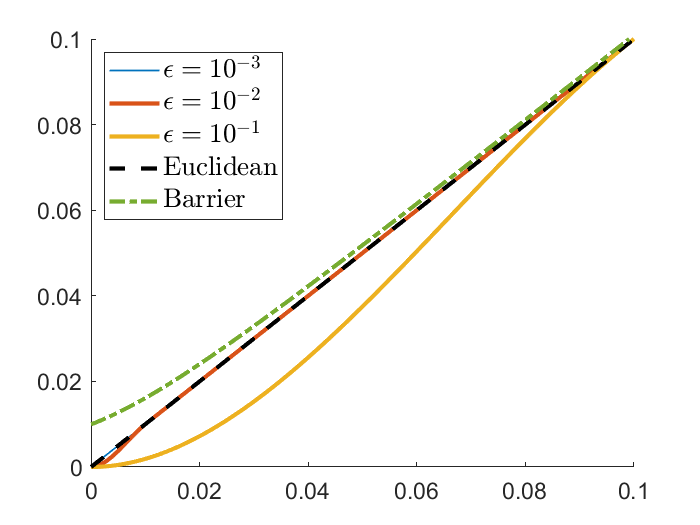}
    \caption{Projection-like function: plot of the function \eqref{proj} for different choices of the parameters $\epsilon$. The dashed line represents the Euclidean projector, while the dash-dot line is for the interior barrier projection proposed in \cite{Bertocchi-etal-2019} with parameter $10^{-4}$.}
    \label{fig:proj}
\end{figure}

The function in \eqref{proj} is, at the best of our knowledge, new. In practice, it is a smooth function which coincides with the Euclidean projector except in $[0,\epsilon]$, $[1-\epsilon,1]$, where it is defined as a third degree polynomial which interpolates the points $(0,0),(\epsilon,\epsilon)$ and $(1-\epsilon,1-\epsilon),(1,1)$, respectively. If compared to the interior barrier function in \cite{Bertocchi-etal-2019}, the projection-like function \eqref{proj} is such that all points outside the feasible region are set exactly equal to the closest bound of the box: we believe that this makes it more complying with the term \eqref{bimodal}. On the other hand, it presents two oscillations close to the interpolation points. The plot of the projection-like function \eqref{proj} restricted to an interval close to the origin is presented in Figure \ref{fig:proj}, for different choices of the parameter $\epsilon$. The plot of the Euclidean projector and of the interior projection function proposed in \cite{Bertocchi-etal-2019} are also reported for further reference. In the numerical experiments presented in Section \ref{sec_tests}, we set $\epsilon = 10^{-4}$.

The variational deconvolution procedure based on the FISTA iteration and on the projection-like function \eqref{proj} is outlined in Algorithm \ref{unrolling}.
\begin{algorithm}
\caption{FISTA-like deblurring procedure}\label{unrolling}
\textsc{Input}: $f\in\R^n$.\\
%\begin{equation}\nonumber%\label{unrolling}
$
\begin{array}{ll}
u^{(0)}   = u^{(-1)}  = f\\
%u^{(0)}   = f\\
\mbox{\textsc{For} }  k = 0,...,K\\
           \left\lfloor\begin{array}{lcl}
						  \bar v^{(k)} &=& u^{(k)} + \beta_k(u^{(k)}-u^{(k-1)})\\[0.2cm]
							 v^{(k)}     &=& \Pi(\bar v^{(k)})\\[0.2cm]
							 t^{(k)}     &=& v^{(k)}-\alpha_k \nabla_u E_f(v^{(k)},\theta)\\[0.2cm]
							u^{(k+1)}    &=& \Pi(t^{(k)})
					    \end{array}\right.\\
u^* = u^{(K+1)}
\end{array}
$\\
%\end{equation}
\textsc{Output}: $u^*$.
\end{algorithm}
In order to mimic the original FISTA algorithm for convex optimization \cite{Beck-Teboulle-2009b}, we choose $\beta_k = (k-1)/(k+1)$, which, as proved in \cite{Chambolle-Dossal-2015,Nesterov-2005}, produce an acceleration of the objective function decrease with respect to standard gradient methods. As concerns as the steplength parameters $\alpha_k$, we adopt the quite common strategy of including them in the set of parameters to be learned, besides the energy functional parameters $\theta$. Then, denoting by $\alpha\in\R^K$ the vector containing all the steplength parameters, we can make explicit these dependencies by writing 
\begin{equation*}
    u^* = u^*(\theta,\alpha).
\end{equation*}

The idea of including the FISTA iteration in a learning procedure has been already proposed in several contexts, often in combination with neural networks and deep learning techniques (see for example \cite{Arridge-etal-2019,Gregor-LeCun-2010,Xiang-etal-2021} and references therein). As explained above, here we consider the FISTA unrolling in the framework of bilevel optimization, which is closer to the traditional variational approach and lead to more explainable results, with far fewer parameters to learn and a consequently reduced computational cost. 

The computation of the gradient of the merit function at the upper level in \eqref{prob:unrolling} can be obtained without explicitly forming the Jacobian of the map $u^{(k+1)}(\theta,\alpha): \R^p\to \R^n$, by implementing a backward differentiation algorithm similar to that proposed in \cite[Algorithm 1]{Ochs-etal-2015}. The main steps of this procedure are outlined in Algorithm \ref{loss_gradient} (see Appendix A). 

\section{Learning model}\label{sec_loss}

We are now ready to state our parameter learning problem in a more complete manner: for simplicity, let us assume here and in the following that the training set is composed by only one sample(the generalization can be easily obtained by averaging with respect to all samples). Then, we have to solve
\begin{equation}\label{prob_learn_noreg}
\min_{\theta\in\R^4,\alpha\in\R^K} \ell(u^*(\theta,\alpha)),
\end{equation}
where $\ell$ is a merit function which measures the quality of the reconstruction $u^*$.

\silviacorr{Solving problem \eqref{prob_learn_noreg} will produce the optimal value of the energy functional parameters in \eqref{thetadef} and steplength $\alpha$ of Algorithm \ref{unrolling}.}

In this section we will discuss the choice of the merit function, with specific reference to the HDC problem. This is a crucial issue for the performance of the overall methodology. From one hand, it has to be selected in such a way that the learned image restoration procedure provides output images as much as possible cleaned from blur and noise and, possibly, highly scored by the OCR software. On the other hand, it must be smooth and simple enough to be handled by an optimization method in the parameters learning stage.

In the following, we describe two alternative approaches to this issue: the first one consists in an actual loss function, defined upon a measure of similarity of the restored image $u^*$
with respect some ground truth image $g$. This idea is well established in the framework of (deep) learning techniques for image restoration: however, it strongly relies on the availability of a good ground truth image. As we will discuss in the next section, this is not the case of the original set of HDC data. To circumvent this drawback, we propose also another approach which does not make explicit use of a ground truth image (in this sense, it is an unsupervised strategy), but it is based on the prediction of the quality of a given image in terms of its OCR score. In sections \ref{sec:ssim} and \ref{sec:SVR} we outline the main ideas behind these two approaches. 
 
\subsection{Supervised approach: SSIM--based loss function}\label{sec:ssim}

The problem of defining a meaningful image metric is a relevant issue. In the simplest case, it can be expressed by means of the squared Euclidean distance. However, in the context of HDC data, we observed that, since it is based on a pixel-by-pixel evaluation, it may often lead to penalize images with good OCR scores, while promoting images with evident artifacts.\\ 
For this reason we adopt the well known SSIM metric, in the definition of the original paper \cite{Wang-etal-2004}:
\begin{equation}\label{SSIM}
    SSIM(u,g) = \frac 1 m \sum_{i=1}^m S_i^{(1)}(u,g)\cdot S_i^{(2)}(u,g),  \end{equation}
    with
\begin{equation}\label{SSIM2}    S_i^{(1)}(u,g)=\frac{2\mu^u_i\mu^g_i+C_1}{(\mu_i^u)^2 + (\mu_i^g)^2+C_1},\ \ S_i^{(2)}(u,g) = \frac{2\sigma^{ug}_i+C_2}{\sigma_i^u + \sigma_i^g+C_2},
\end{equation}
where $\mu^u,\mu^g, \sigma^u,\sigma^g,\sigma^{ug}\in\R^m$ are defined below. Let us first denote by $W\in\R^{m\times n}$ the convolution matrix corresponding to a given low pass/window kernel selecting only the valid pixels. Then, we set
\begin{eqnarray*}
\mu^u_i = \sum_{j=1}^n W_{ij}u_j, & & \mu^g_i = \sum_{j=1}^n W_{ij}g_j, \\
\sigma_i^u =\sum_{j=1}^n W_{ij}({u}_j)^2 - (\mu^{u}_i)^2, & & \sigma_i^g = \sum_{j=1}^nW_{ij}({g}_j)^2 - (\mu^{g}_i)^2,\\
\sigma_i^{ug} =\sum_{j=1}^n W_{ij}\mu^{u}_j\mu^{g}_j - \mu^{u}_i\mu^{g}_i, & & i = 1,\ldots,m.
\end{eqnarray*}
%the images obtained by convolving the image $u$ (resp. $g$) with a window kernel $w$; $\sigma^u,\sigma^g$ are obtained by convolving the window $w$ with the images whose pixels are those in $u$ (resp. $g$), squared; similarly $\sigma^{ug}$ denotes the convolution between $w$ and the image whose pixels are given by $u_ig_i$. 
As window kernel, we adopt a Gaussian filter with standard deviation $1.5$ and size of $11$ pixels, while the coefficients $C_1, C_2$ are set equal to $10^{-4}$ and $3\cdot 10^{-4}$, respectively (these are standard values for image ranging in $[0,1]$). The SSIM index takes values in $[0,1]$ and scores close to 1 indicate a high degree of similarity between $u$ and $g$.

The function in \eqref{SSIM} is smooth, then we consider the following merit function for the parameters learning phase:
\begin{equation}\label{lossSSIM}
    \ell(u^*(\theta,\alpha)) = 1 - SSIM(u^*(\theta,\alpha),g).
\end{equation}
In particular, its gradient w.r.t. $u$ can be computed by first defining the quantities
\begin{eqnarray*}
    q_i &=& \frac{S_i^{(2)}(u,g)} {(\mu_i^u)^2 + (\mu_i^g)^2+C_1}\left( \mu_i^u-\mu_i^g\frac{2\mu_i^u\mu_i^g+C_1}{(\mu_i^u)^2 + (\mu_i^g)^2+C_1}\right)\\
    t_i &=&\frac{ S_i^{(1)}(u,g)}{\sigma_i^u + \sigma_i^g+C_2}\left(\mu_i^g-\mu_i^u\frac{2\sigma^{ug}+C_2}{\sigma_i^u + \sigma_i^g+C_2}\right)\\
    r_i &=& \frac { S_i^{(1)}(u,g)}{\sigma_i^u + \sigma_i^g+C_2}\\
    z_i&=& S_i^{(2)}(u,g)\cdot r_i,
\end{eqnarray*}
for $i=1,\ldots,m$. Then, for $j=1,\ldots,n$, we have
\begin{equation*}
    [\nabla_u\ell(u)]_j = -\frac{2}{m}\sum_{i=1}^mW_{ij}(q_i-t_i) + \frac{2}{m} g_j\sum_{i=1}^m W_{ij}r_i - \frac{2}{m} u_j\sum_{i=1}^m W_{ij}z_i.
    %\left(W^T(q-t) + GW^Tr- UW^Tz\right),
\end{equation*}

\subsection{Unsupervised approach: SVm for Regression}\label{sec:SVR}

The SSIM--based approach described in the previous section makes heavily use of the ground truth images provided by the first camera, which are anyway still affected by nonuniform background and radial distortion. A different approach might consist in maximizing the OCR function itself for a given input image. However, it is not possible to directly use OCR as a loss function, since it is provided as a black-box. As an alternative, we propose to construct a loss function able to predict the OCR score starting from a training set $\{(x^{(s)},y^{(s)})\}_{s=1}^{S}$ consisting of images and the corresponding OCR values. This prediction function can be obtained by solving a standard regression problem in a supervised context. In order to obtain a differentiable prediction function, we have excluded all the machine learning techniques based on binary decision trees, such as Random Forest. Given the need to create an ad hoc dataset, we also had to exclude Deep Learning methodologies, which would have required an excessive amount of data. The resulting natural choice has been the well-known Support Vector machines for Regression (SVR) methodology \cite{Drucker-etal-1996, Cristianini2000}. Previously in the literature, SVR has been employed as a performace predictor within methods that needed guidance in choosing the hyperparameters of a network \cite{NAS}, similar to how it will be employed in this paper. In this framework, the decision function assigns a predicted label $y_{\text{pred}}$ to a new image $x$ through the following formula
\begin{equation}\label{SVR}
y_{\text{pred}} = F(x) = \sum_{s=1}^{S} (\bar{\upsilon}_s-\bar{\upsilon}_s^*)K(x^{(s)},x)+ \bar{b},
\end{equation}
where $K:\mathbb{R}^n \times \mathbb{R}^n \longrightarrow \mathbb{R}$ is a kernel function \cite{Scholkopf2002}, $\bar{\upsilon},\bar{\upsilon}^*\in\mathbb{R}^S$ are computed as the solution of the constrained quadratic programming problem
\begin{eqnarray}\label{SVReq1}
\min_{\upsilon, \upsilon^* \in \mathbb{R}^S}  && \frac{1}{2} \sum_{s,s'=1}^S (\upsilon_s - \upsilon_s^*)(\upsilon_{s'}-\upsilon_{s'}^*) K(x^{(s)},x^{(s')}) + \varepsilon \sum_{s=1}^S (\upsilon_s+\upsilon_s^*) - \sum_{s=1}^S y^{(s)}(\upsilon_s-\upsilon_s^*) \nonumber\\
\text{s.t.} &&  \sum_{s=1}^{S} (\upsilon_s - \upsilon_s^*) = 0 \nonumber\\
&&  0 \leq \upsilon_s \leq C, \quad 0 \leq \upsilon_s^* \leq C, \qquad s=1,\ldots,S
\end{eqnarray}
and $\bar{b}$ is determined by one of the two relations
\begin{align}\label{SVReq2}
\bar{b} &= y^{(s')}-\sum_{s=1}^{S} (\bar{\upsilon}_s-\bar{\upsilon}_s^*)K(x^{(s)},x^{(s')})-\varepsilon\nonumber\\
& \\[-5mm]
\bar{b} &= y^{(s')}-\sum_{s=1}^{S} (\bar{\upsilon}_s-\bar{\upsilon}_s^*)K(x^{(s)},x^{(s')})+\varepsilon\nonumber
\end{align}
for a given training sample $(x^{(s')},y^{(s')})$ for which either $0 < \bar{\upsilon}_{s'} < C$ or $0 < \bar{\upsilon}_{s'}^* < C$.\\
The two hyperparameters $C$ and $\varepsilon$ in \eqref{SVReq1}--\eqref{SVReq2} handle the bias-variance dilemma, since they represent an upper bound for the components of the coefficients $\upsilon_s,\upsilon_s^*$ and the width of the insensitive zone in the SVR loss function, respectively \cite{Cristianini2000}. As concerns the kernel function, in our experiments we chose the Gaussian kernel defined as
\begin{equation*}\label{Gauss_ker}
K(x,x')=e^{ \displaystyle -\dfrac{\|x-x'\|^2}{2\sigma^2}}, \qquad \sigma > 0.
\end{equation*}

By exploiting the SVR prediction function, we can define the following merit function:
\begin{equation}\label{lossSVR}
\ell(u^*(\theta,\alpha)) = \displaystyle e^{- \dfrac{F(u^*(\theta,\alpha))}{100}},
\end{equation}
where $F$ is defined in \eqref{SVR} and the division by $100$ has the meaning of normalisation, the OCR values being between $0$ and $100$. This function must then be derived with the chain rule in order to optimise against the parameters. We present the derivation in $u$ in the following formula:

\begin{equation}
\scalebox{0.975}{$
   \nabla_u \ell(u^*(\theta,\alpha))= \frac{1}{100}e^{- \frac{F(u^*(\theta,\alpha))}{100}}
   \displaystyle \sum_{s=1}^{S} (\bar{\upsilon_s}-\bar{\upsilon_s}^*)e^ -\frac{\|u^*(\theta,\alpha)-x^{(s)}\|^2}{2\sigma^2}\left( \frac{u^*(\theta,\alpha)-x^{(s)}}{\sigma^2} \right)$}
\end{equation}

\begin{comment}
\begin{equation}
   \nabla_u \ell(u^*(\theta,\alpha))=e^{- \dfrac{F(u^*(\theta,\alpha))}{100}}\sum_{s=1}^{S} (\bar{\upsilon_s}-\bar{\upsilon_s}^*)e^{ \displaystyle -\frac{\|u^*(\theta,\alpha)-x^{(s)}\|^2}{2\sigma^2}}\left( \frac{u^*(\theta,\alpha)-x^{(s)}}{100\sigma^2} \right)
\end{equation}
\end{comment}

An interesting point to emphasise in the case just presented is the fact that, whereas for the SSIM loss, argued in the previous section, we need the ground truth of the training images, in this case it is not necessary, making the approach particularly interesting in real-world contexts, where the true images are rarely known.

\section{Numerical experiments}\label{sec_tests}

In this section we provide more details on the bilevel optimization model, as well as the results we obtained using the two different loss functions. In Table \ref{tab:hdc_results} we have reported the official results of the challenge obtained for the steps 6, 8 , 10 and 12, with step 6 being the first step we failed to pass. Our goal for this work was to improve the model originally submitted and, consequently, the performance.

\begin{table}[!ht]
\centering
\caption{Results of the HDC published in November 2021, with our original placement. We are team number 04.}
\label{tab:hdc_results}
\begin{tabular}{c | c c c c|c}
    \toprule
    \textbf{Team} & \textbf{Step 6} & \textbf{Step 8} & \textbf{Step 10} &  \textbf{Step 12} &\textbf{$\#$parameters} \\
    \midrule \midrule
    15\_A & 94.03 & 93.12 & 93.75 & 91.42&$2.6$ millions \\
    \midrule
    12\_B & 92.62 & 92.62 & 85.80 & 85.95&$11$ millions\\
    \midrule
    01 & 91.75 & 91.65 & 88.67 &87.12 &$0.187$ millions\\
    \midrule
    11\_C & 87.78 & 81.25 & 79.15 &62.80 & $52$ millions \\
    \midrule
    06 & 94.33 & 85.92 & 70.17 &0.00 &$2.6$ millions\\
    \midrule
    13 & 71.12 & 67.12 & 54.38 &	64.83 &$2.2$ millions\\
    \midrule
    16\_B & 76.45 & 68.35 & 4.03 &7.42 &3 \\
    \midrule
    \textbf{04} & \textbf{68} & \textbf{62.85} & \textbf{24.38} &\textbf{10.70} &\textbf{4} \\ 
    \midrule
    09\_B & 6.33 & 2.27 & 2.62&4.03 &4  \\
    \bottomrule
\end{tabular}
\end{table}

Referring to Table \ref{tab:hdc_results}, and without claiming to be either precise or exhaustive, we report a quick analysis of the methods proposed by the other teams, with particular attention to the number of parameters trained and the proposed techniques. 
\begin{enumerate}
    \item 15\_A, Technische Universität Berlin, Institut für Mathematik, Berlin, Germany, proposed a end-to-end deblurring neural network, whose architecture is a slight modification of the standard U-Net \cite{UNet}, with about $2.6$ million parameters. Some tricks to avoid overfitting and make the method generic were incorporated.%, while still losing the interpretability of the model due to the use of deep learning techniques. 
    \item 12\_B, Institution Department of Mathematics, National University of Singapore, proposed a deep learning approach borrowed from the DeblurGanV2 \cite{DGV2}, without the use of the GAN training loss.
    \item 01, Leiden University, Leiden, The Netherlands, used Mixed-Scale Dense CNNs \cite{dense} to deconvolve images of text.
    \item 11\_C, ZeTeM Uni Bremen Team, used a fully-learned and purely data-driven inversion model, the StepNet itself consists of 20 sub-networks which are connected in sequence. Each StepNet receives an input with blurring level $i$ and produce an output with blurring level $i-1$. For the implementation of the StepNet model, the author uses 20 small U-Nets \cite{UNet} for the sub-networks.
    \item 06, University Düsseldorf, Department of Computer Science Germany, after data augmentation with DIV2K dataset, trains a neural network to deblur images from both the DIV2K dataset and the HDC2021 dataset. The neural network used here is adapted from \cite{FB} and uses a UNet architecture (see \cite{Germer-etal-2022}).
    \item 13, Federal University of ABC; Center for Engineering, Modeling and Applied Social Sciences - Brazil, uses as the main idea the Deep Image Prior (DIP) \cite{Ulyanov-etal-2020} reconstruction, which uses only the degraded image. Instead of using the DIP alone, a second DNN with bottleneck architecture (as an autoencoder) is used to help the deblurring task, as it includes (prior) information from the sharp images too. 
    \item 16\_B, Technical University of Denmark, DTU Compute Denmark, implements an image deblurring algorithm with Point-Spread-Function (PSF) radius estimation \cite{Nicolai}.
    \item 4, our team, 4 manually tuned parameters of a simple variational model involving smooth TV regularization plus non-negativity constraints and solved via a gradient projection method.
    \item 09\_B, University of Campinas (UNICAMP), School of Electrical and Computer Engineering - Brazil, propose a Regularization by Denoising \cite{Brazil} method. First they estimate the PSF with the given dot images at each step and then perform deconvolution using an inverse-problem framework with the RED (Regularization by Denoising) fixed point algorithm.
\end{enumerate}
 
We observe that most competitors employed convolutional neural networks, i.e., deep learning techniques, by adapting the best-known architecture with fewer parameters with respect to standard implementations, while only Teams 16\_B, 4, and 09\_B proposed traditional variational methods. The challenge results, summarized in Table \ref{tab:hdc_results}, show that the DNN achieve much better results. Actually, Team 15\_A and 12\_B were able to pass also steps 18 and 19 of the challenge, while only Team 16\_B was able to pass level 6 without the use of neural networks, even if they obtained very bad performances on step 10.

\subsection{Dataset definition for SSIM optimization}\label{sec:dataset_ssim}

In this section we briefly describe the construction of the dataset to be employed in the supervised case, i.e., a set of samples $\{(f^{(s)},g^{(s)})\}_{s=1}^S$, where $f^{(s)}$ is a blurred, possibly noisy, version of the ground truth image $g^{(s)}$. Actually, the images provided in the HDC dataset can not be directly employed for this purpose. Indeed, the images detected by CAM2 are not only blurred, but also contain a nonuniform background and suffer of optical distortions. For these reasons, we perform a preprocessing of the HDC data for (partly) removing these two perturbation effects and build the training set for the SSIM based merit function.\\
In the dataset formation and also in the subsequent learning and restoration procedure, we consider a flipped version of the images from CAM1 and CAM2 in the original HDC dataset, to have white characters over black background. In order to speed up the computations, we also reduce the dimension of the images up to a factor of $1/8$ and rescale the pixel values so that they range in $[0,1]$. \\
After these preliminary operations, for each blur level, we first estimate the background by considering the average of the images from CAM2 on the frame around the writings area and defining the pixel values in the central part by interpolation. Once obtained this estimation, we subtract it from all images to define the response image $f^{(s)}$. An example of the estimated background for the 10--th blur level is shown in Figure \ref{fig:dataset_ssim} (c) and (d).

As for the optical distortion, we adopt a quite simple radial model with only two parameters, which have been manually tuned to construct a ground truth matching with the response image. In order to give some more details about this procedure, let us denote by $g_{CAM1}$ the binarized version of one image from CAM1, for a given blur level. Let us introduce also the notation $(x_i,y_j)$, $i=1,...,n_1$, $j=1,...,n_2$ for the spatial coordinates of the gridpoints corresponding to the image pixels. 
Then, we consider an interpolation function $g:\R^2\to\R$ such that 
\begin{equation*}
    g(\bar{x}_i,\bar y_j) = [g_{CAM1}]_{ij}, \ \mbox{ where }
    \left\{\begin{array}{lcl}\bar x_i &=& c_x + R(x_i-c_x) \\
    \bar y_j &=& c_y + R(y_j-c_y)
    \end{array}\right.,
\end{equation*}
where $R>0$ and $(c_x,c_y)$ are the radius and the center of the distortion. With a little abuse of notation, we still denote by $g$ also the image obtained by sampling the interpolation function over the gridpoints.

In order to determine the parameters $R,c_x,c_y$, we first compute an acceptable restoration of the background subtracted image from CAM2 corresponding to $g_{CAM1}$, with a simple TV based variational method. Then, we compute the binarization of the restored image, which will be denoted by $\hat u$. Finally, $R,c_x,c_y$ have been manually tuned to have a good match between the edges of $\hat u$ and those of $g$ (see Figure \ref{fig:dataset_ssim} (e) and (f)). Even if the optical distortion could be included in the model and its parameters learned as well as the ones connected to the variational procedure, we prefer to perform the correction directly on the dataset in order avoid a further nonlinearity and to preserve the primary aim of the challenge, which is focused on deblurring. Moreover, we choose to correct the dataset by warping the ground truth instead of unwarping the data, to avoid introducing an additional perturbation on them.

To summarize, in each pair $(g^{(s)},f^{(s)})$ of the training set for the SSIM loss function, the image $f^{(s)}$ is obtained by subtracting the estimated background from the CAM2 data, flipped and resized, while $g^{(s)}$ is obtained
by applying the estimated radial distortion to the binarization of the corresponding image from CAM1.
\def\scalfact{0.55}
\begin{figure}
    \centering
    \begin{tabular}{cc}
    \includegraphics[scale=\scalfact]{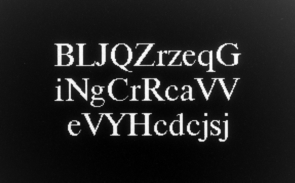} &  \includegraphics[scale=\scalfact]{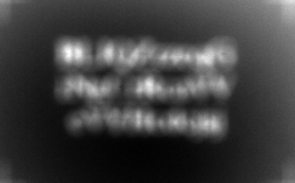}\\
    (a)&(b)\\
    \includegraphics[scale=\scalfact]{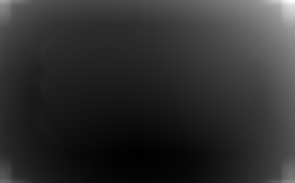} &  \includegraphics[scale=0.35]{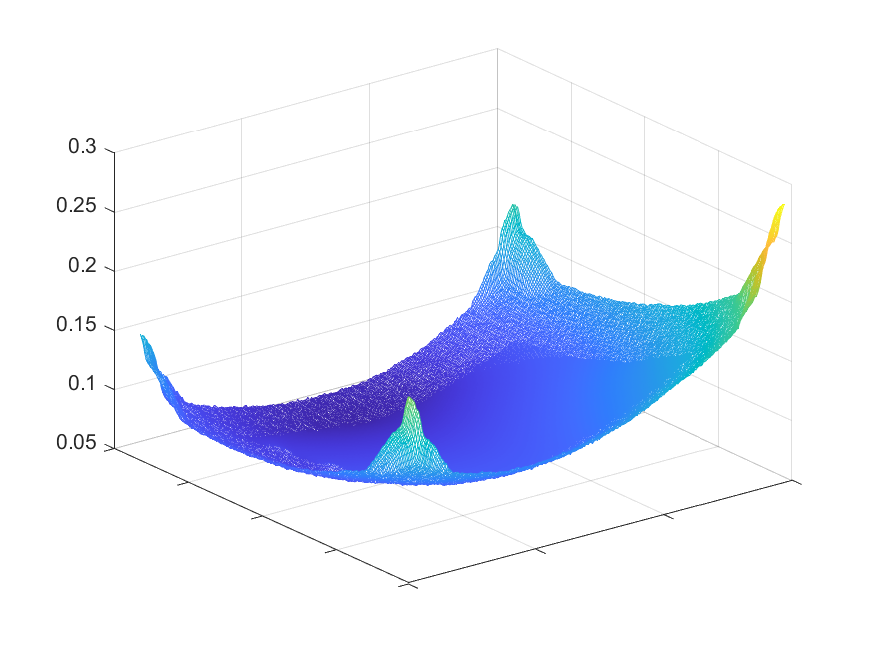}\\
    (c) & (d)\\
    \includegraphics[scale=\scalfact]{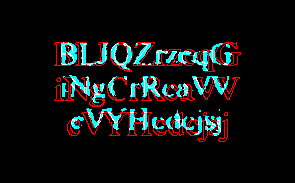} &  \includegraphics[scale=\scalfact]{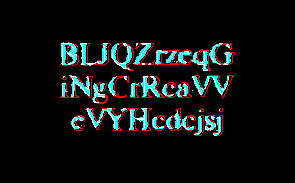}\\
    (e) & (f)
    \end{tabular}
    \caption{Dataset definition for the SSIM loss function. Upper row: original HDC data from step 10, Times New Roman, image n.51, CAM1 (a) and CAM2 (b). Middle row: estimated background, represented as an image (c) and as a surface (d). Bottom row: estimated distortion. Panel (e) shows the binarized reconstruction (light blue) of image (b) obtained with a variational method and the edges of the binarized version of (a) (red). The white lines correspond to the pixels where the two images are superimposed. In panel (f), the red edges are obtained after applying the estimated radial distortion to (a).    }
    \label{fig:dataset_ssim}
\end{figure}

\subsection{OCR predictor via Support-Vector machine for Regression}\label{sec:dataset_SVR}

%The other approach we followed was to predict the OCR score with the SVR algorithm and use this information to build a merit function that never takes into consideration the ground truth of the image in any shape or form, not even its score, making this an unsupervised learning approach. Here we present more rigorously how we implemented the SVR loss function and the outcome of the training of the energy functional. We also discuss in detail the dataset for the SVR training.
In this section we describe the procedure we define for training the SVR to predict the OCR score of a given image. Due to its data driven nature, the SVR requires a good dataset to achieve significant results, i.e., the collected examples should be numerous, representative and comprehensive for the addressed problem.

For our specific HDC application, we created a mixed real/synthetic training set of images, mainly due to the fact that the score assigned by the OCR to the HDC images was, in the vast majority of cases, either above 60 or just 0, thus resulting in an unbalanced dataset. In particular, we selected 400 samples from the ones given by the challenge organizers, corresponding to 50 corrupted images for each step from 1 through 8. Moreover, we generated thousands of images with three lines of characters (including numbers) similar to the ground truths of the HDC, and we created 8 blurred versions of each one, by convolving them with a circular PSF of different radius. In order to create a well distributed dataset, we selected 123 images for each score interval $[0, 10], ]10,20], \ldots, ]90,100]$, for a total of 1230 supplemental examples. We simulated the images which would result by applying Algorithm \ref{unrolling} with wrong choices of the parameters. With this idea, we also included constant images, with and without the addition of some low variance noise, to penalize these kind of samples in the eyes of the SVR. Moreover, we actually applied the unrolled optimization algorithm with random, mostly wrong, parameters to the images of the HDC set and then we included the reconstructions in the dataset, for an additional 429 images. An overestimation of the radius of the PSF leads to a ringing effect which is not only unpleasant to the eye, but also a potential hindrance to the OCR. Thus, the inclusion of these other extra samples is required in order to diversify the dataset and increase the reliability of the SVR when learning the energy functional. 

\begin{figure}[!ht]
  \centering
  \begin{subfigure}{.3\linewidth}
    \centering
    \includegraphics[width = \linewidth]{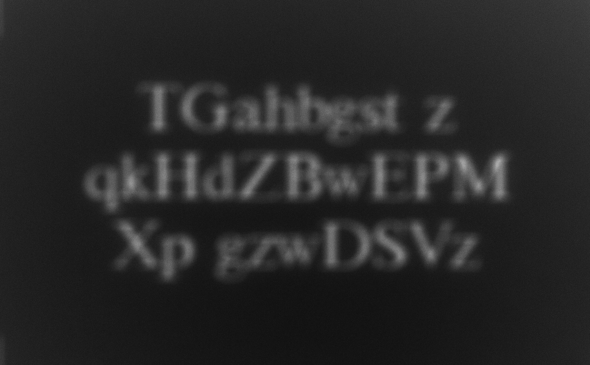}
    \caption{}
  \end{subfigure}%
  \hspace{.2cm} % Space between image A and B
  \begin{subfigure}{.3\linewidth}
    \centering
    \includegraphics[width = \linewidth]{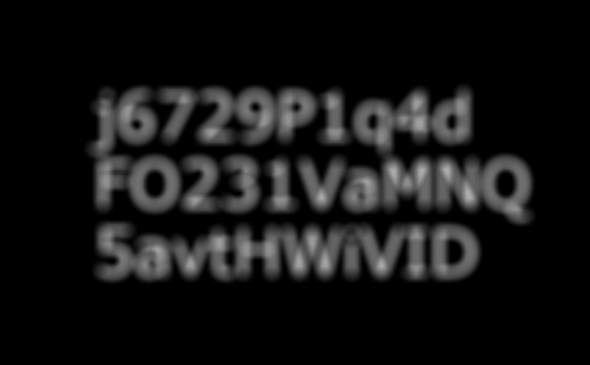}
    \caption{}
  \end{subfigure}%
  \hspace{.2cm}% Space between image B and C
  \begin{subfigure}{.3\linewidth}
    \centering
    \includegraphics[width = \linewidth]{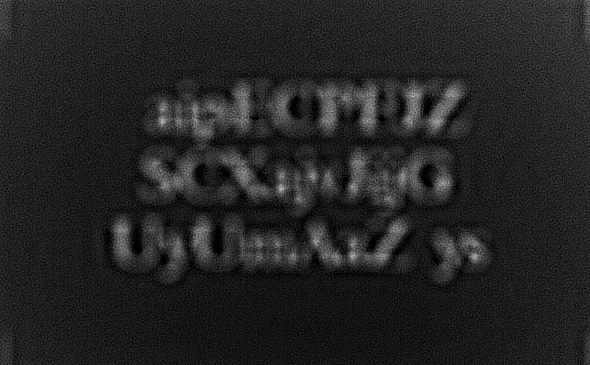}
    \caption{}
  \end{subfigure}
  \caption{Samples from the SVR training dataset. (A) Noisy image from the step 4 of the Helsinki dataset (OCR score = 52); (B) Synthetic image blurred with a disc PSF of radius 9 (OCR score = 84); (C) Reconstruction with wrong parameters for the variational model (OCR score = 11).}
  \label{fig:svr_samples}
\end{figure}

To improve the performance of the SVR, all the images were resized by a factor of $1/4$, to have dimensions of $365 \times 590$, and flipped to have white text on a black background. Just before their evaluation by means of OCR they are flipped again and restored to the original size. Figure \ref{fig:svr_samples} shows one sample for each kind. As expected, the synthetic image, with no noise or distortion, has the best score of 84 out of the three, while the third image is an example of how the ringing effect is detrimental for the performance by reaching a score of just 11.

Regarding the SVR hyperparameters mentioned in Section \ref{sec:SVR}, finetuning work was done, using cross validation techniques to have a wide generalization capability. The final values used to train the predictor are $\varepsilon=4.8$ and $C=48$, obtaining an average error value on the test set of $16$ OCR points.

\vskip 3mm
\subsection{Setting of the bilevel problem}

In this section we give some details about the setting of problem \ref{prob:unrolling} and its hyperparameters, assuming that one of the two merit functions defined in Sections \ref{sec:ssim} and \ref{sec:SVR} is employed, equipped with the proper dataset described in Sections \ref{sec:dataset_ssim} and \ref{sec:dataset_SVR}.

We tested the model with three different numbers of inner iterations, more specifically with $K = 60,70,80$. When using a bilevel model, it is fundamental that the reconstruction $u^*$ is as close as possible to a minimum of the energy functional. For this reason, we expected that fewer inner iterations were not enough for the higher steps of the challenge, since stronger blur requires more iterations of the optimization algorithm in the classical variational approach to produce a good quality restoration. 

%The parameters learned were $\theta = (r,\rho,\gamma,\epsilon)$, as well as the steplength sequence $\{\alpha_k \}_{k=0}^K$ of the unrolled algorithm. Moreover $\gamma = \gamma_1 = \gamma_2$, which means that the TV-like regularizer is isotropic. Also, since most pixels in the reconstructed images would not reach the maximum value one making the text characters in the image a bit "foggy", we slightly modified the second term in the energy functional, that now is formulated as:
%\[
%E_f(u,\theta) = \frac 1 2 \|H(r)u - f\|^2 + \frac \rho 2 \sum_{i=1}^n %u_i(0.6-u_i) +  \gamma \sum_{i=1}^n\phi(\nabla_i^{h}u) + \phi(\nabla_i^{v}u)
%+ \iota_{[0,1]}(u)
%\]

Regarding the upper level problem, all the parameters of interest are bounded within reasonable intervals, which have been easily determined when looking for a good starting point. This also serves as a form of regularization, preventing the loss function from overfitting and also from introducing artifacts due to noise and the incomplete knowledge of the blurring kernel. The upper level problem was numerically solved by using the Scaled Gradient Projection method \cite{Bonettini2009,Bonettini2015,Bonettini-etal-2019} until a relative difference of $10^{-7}$ between the merit function value at two successive iterates or a maximum of 50 iterations was reached. To accommodate the different scalings of the components of the loss gradient, different strategies for the stepsize selection were imposed, specifically for each entry of the optimization variable $\theta$. 
%In table \ref{tab:svr_bounds} we reported the bound for each parameter. Only the radius has a variable upper limit since, as explained before, a too high radius would result in a ringing effect which is not totally understood by the SVR.

%\begin{table}[!ht]
%\centering
%\caption{Summary of the lower and upper bounds for the parameters of the %variational model. Specifically, $\alpha$ denotes the generic steplength of %$\mathcal{A}$.}
%\label{tab:svr_bounds}
%\begin{tabular}{c | c c c c c c c}
%\toprule
%            & $\gamma$ & $\rho$ & $\delta$ & $\alpha$ & $r$ step 6 & $r$ step 8 & $r$ step 10  \\
%            \midrule \midrule
% Min. value & $10^{-6}$ & $10^{-5}$ & $10^{-4}$ & $10^{-8}$ & $5$ & $5$  & $5$ \\
%  \midrule
% Max. value & $10^{-3}$ & $3 \times 10^{-3}$ & $5 \times 10^{-1}$ & $2$ & $11$ & $15.499$  & $20$ \\
% \bottomrule
%\end{tabular}
%\end{table}

As for the computational cost, the training phase was done using only $4$ noisy samples from the ones provided by the HDC, all with the font Times (the more complicated of the two), for both the merit functions \eqref{SSIM} and \eqref{SVR}. Another key difference with respect to deep neural networks is the lesser need of samples in the parameter learning phase. In this way, we managed to balance the computationally expensive evaluation of the merit function and its gradient, keeping the training time in a reasonable bound. In particular, the solution of the bilevel problem with the SSIM based merit function \eqref{SSIM} takes about 15-30 minutes on a workstation equipped with a multicore CPU Intel(R) Core(TM) i7-6700 CPU @3.40GHz, while the solution time with the SVR merit function \eqref{SVR} requires few hours on the same architecture. All the routines have been implemented and run in Matlab R2021b.

%Lastly, after flipping the images again to have black text over white background, we post-processed them by fixing to 1 the value of all the outer pixels. This was done for two reasons. The first one was to avoid having the reconstructions ruined by the normalization of the OCR. On the other hand, the points of reference used to take the pictures were partly present in the images due to them not being perfectly centered. Thus in some reconstructions there would be artifacts at the corners which would be picked up by the OCR as text.

\vskip 3mm

\subsection{Results}

The trained model was applied to the set of text images of steps 6, 8, 10 and 12. \silviacorr{Actually, the HDC test set includes also other kinds of images, with the purpose of checking that the competing algorithms were able to really act as deconvolution methods. For completeness, we report in Appendix B the results obtained by our algorithm to this subset of the test set, together with some remarks about the possible generalization of our approach.}  All reconstructions from the test set were computed by applying Algorithm \ref{unrolling} with the learned parameters to the resized, background subtracted images recorded by CAM2 (the background estimation was obtained as described in Section \ref{sec:dataset_ssim}). 

\silviacorr{As for the computational cost of the reconstruction task, once the parameters have been learned, the restoration of a single image simply requires the computation of $K$ iterations of Algorithm \ref{unrolling}. Then, the overall computational time of the deconvolution procedure is increasing with $K$. We recall that, within the settings described in the previous sections, we have a preprocessing phase of the input image, which is resized to $1/4$ or $1/8$ of its original size for the SVR and SSIM based approach, respectively. This implies that the time needed by the SVR based approach is larger than that corresponding to the SSIM approach, even if the number of unrolled iteration is the same. In particular, the SVR based restoration of a single image takes 1.15, 1.29 or 1.57 seconds (average times over 5 runs) for $K=60,70,80$, respectively, while for the SSIM approach the computational time is 0.37, 0.44 or 0.50 seconds. In all cases, the image restoration requires from a fraction of to little more than a second.}

In Table \ref{tab:newresults} we reported the final average scores obtained on the actual test sets of the challenge. Comparing these results against Table \ref{tab:hdc_results}, we can observe that we achieve an overall performance which is quite close to that of DNN methods, with an approach that can be set also in an unsupervised manner, requires a limited amount of easily recovered examples, is clear and interpretable and depends on at most $84$ parameters instead of millions. It can be also noticed that our mixed bilevel-- unrolling approach can improve the purely variational ones, as those adopted by teams 16\_B, 04, 09\_B.\\

\begin{table}[!ht]
\centering
\caption{Average OCR scores obtained on the 40 test images used in Table \ref{tab:hdc_results}, with the merit functions \eqref{lossSSIM} and \eqref{lossSVR} (columns SSIM and SVR, respectively)}\label{tab:newresults}
\centering
\begin{tabular}{c | c c | c c | c c}
\toprule
        & \multicolumn{2}{c|}{$K = 60$} & \multicolumn{2}{c|}{$K = 70$} & \multicolumn{2}{c}{$K = 80$} \\
        \cline{2-7}
                & SSIM & SVR   & SSIM   & SVR    & SSIM & SVR \\
\midrule
        Step 6  &  85.20  & 85.60  &  85.60& 82.45   & 85.08& 83.28 \\ 
\midrule
        Step 8  & 83.88 & 82.63    & 84.15    & 81.80   & 82.45 & 80.13 \\ 
\midrule 
        Step 10 & 70.88 & 73.90    &  71.35  & 76.30   & 72.72 & 73.23 \\ 
\midrule
        Step 12 &60.23 & 61.53 &61.73 & 48.58 &61.90 & 61.53 \\
\bottomrule
\end{tabular}
\end{table}

A key observation, which is not surprising, is that while the model maintains a similar quality of reconstruction between the steps 6 and 8, the same is not true for the steps 10 \silviacorr{ and 12}. This is most definitely due to a limitation of the variational model at the lower level, which here is still on the simpler side. Indeed, the increasing amount of noise, combined with a good, but not exact, estimation of the PSF, makes these results close to the possible ceiling. 

Regarding the three different values of inner iterations $K$, a bit surprisingly, there is not a clear cut best choice for all blur levels. %Despite the theory telling us that a stronger blurring requires more iterations to be removed, we can see it is not the case here. 
A possible explanation can be found in the fact that the steplengths in Algorithm \ref{unrolling} are learned, probably making up for the lack of iterations. 
%Finally, it is interesting to notice that only step 8 seems to be affected by overfitting, as the average score on the test set is lower by a considerable margin than that on the train set. The same is not true for the other two levels, where the score improves, although with variable magnitude. Such increase is motivated by the presence of the Verdana font in the test set and its absence from the train set. Compared to the Times font, it is less complex and thus its characters are more recognizable.

\begin{figure}[!ht] %% Volendo per le ricostruzioni c'è la versione con le bounding box dell'OCR nella cartella figures, ma si vedono poco.
  \centering
  \begin{subfigure}{.3\linewidth}
    \centering
    \includegraphics[width = \linewidth]{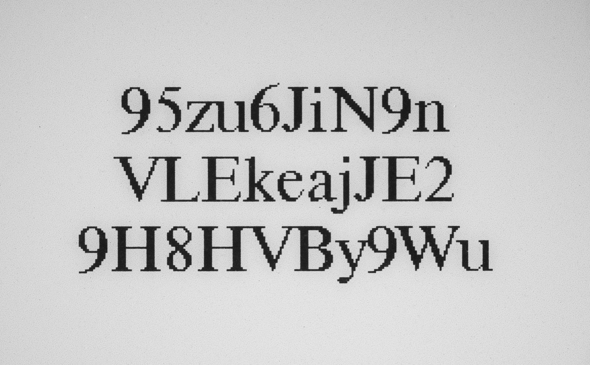}
    %\caption{}
  \end{subfigure}%
  \hspace{.2cm} % Space between image A and B
  \begin{subfigure}{.3\linewidth}
    \centering
    \includegraphics[width = \linewidth]{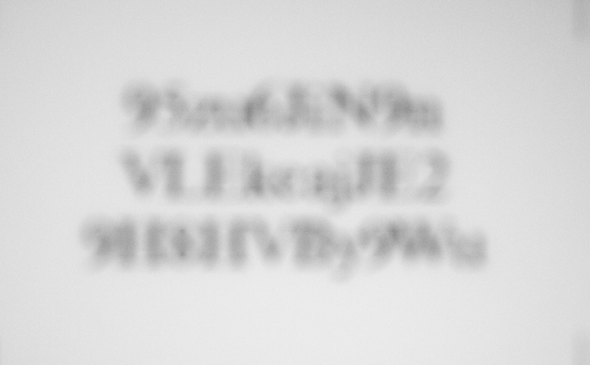}
    %\caption{}
  \end{subfigure}%
  \hspace{.2cm}% Space between image B and C
  \begin{subfigure}{.3\linewidth}
    \centering
    \includegraphics[width = \linewidth]{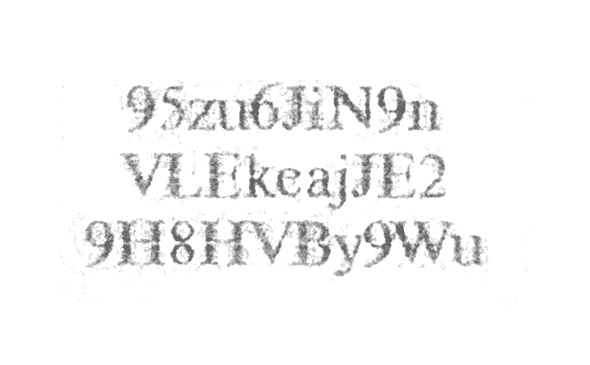}
    %\caption{}
  \end{subfigure} \\
  \vspace{.2cm}
    \begin{subfigure}{.3\linewidth}
    \centering
    \includegraphics[width = \linewidth]{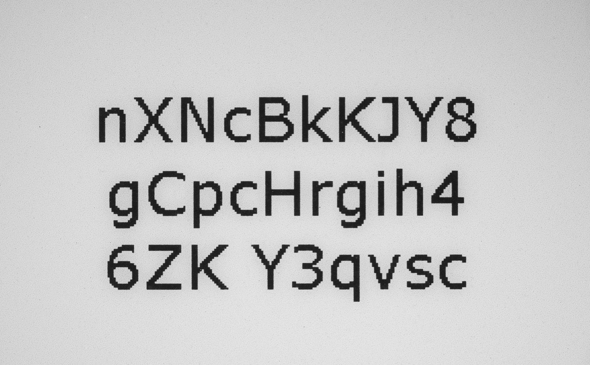}
    %\caption{}
  \end{subfigure}%
  \hspace{.2cm} % Space between image A and B
  \begin{subfigure}{.3\linewidth}
    \centering
    \includegraphics[width = \linewidth]{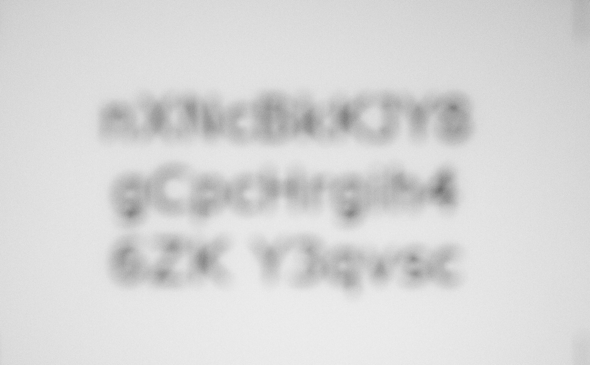}
    %\caption{}
  \end{subfigure}%
  \hspace{.2cm}% Space between image B and C
  \begin{subfigure}{.3\linewidth}
    \centering
    \includegraphics[width = \linewidth]{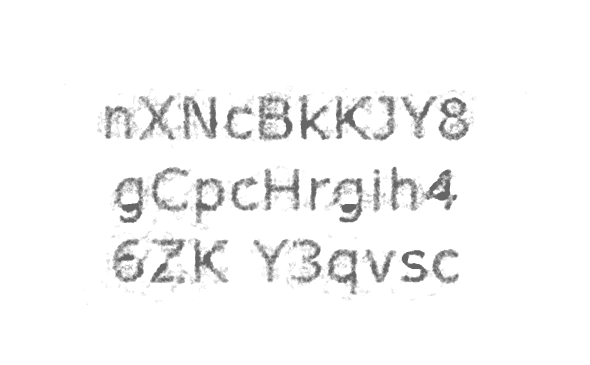}
    %\caption{}
  \end{subfigure} \\
  \vspace{.2cm}
  \begin{subfigure}{.3\linewidth}
    \centering
    \includegraphics[width = \linewidth]{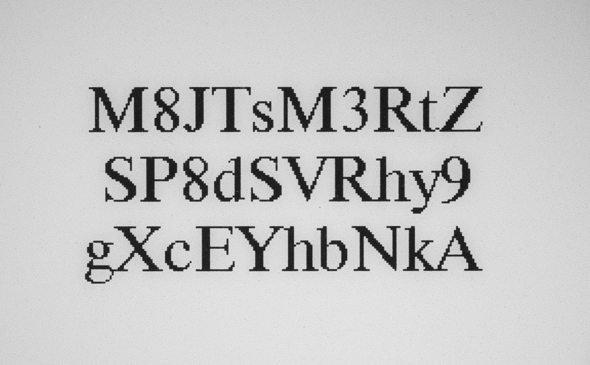}
    %\caption{}
  \end{subfigure}%
  \hspace{.2cm} % Space between image A and B
  \begin{subfigure}{.3\linewidth}
    \centering
    \includegraphics[width = \linewidth]{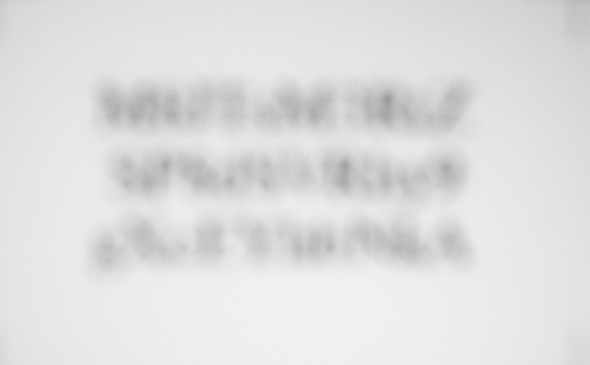}
    %\caption{}
  \end{subfigure}%
  \hspace{.2cm}% Space between image B and C
  \begin{subfigure}{.3\linewidth}
    \centering
    \includegraphics[width = \linewidth]{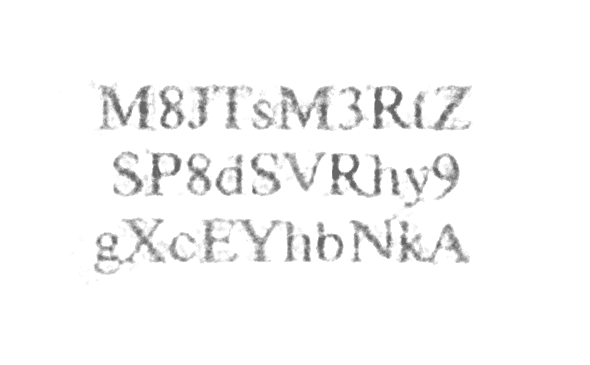}
    %\caption{}
  \end{subfigure} \\
  \vspace{.2cm}
  \begin{subfigure}{.3\linewidth}
    \centering
    \includegraphics[width = \linewidth]{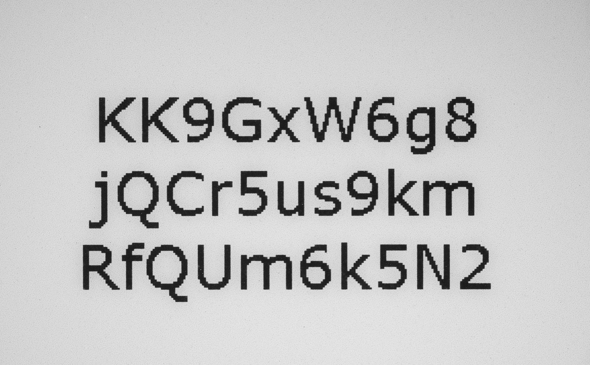}
    %\caption{}
  \end{subfigure}%
  \hspace{.2cm} % Space between image A and B
  \begin{subfigure}{.3\linewidth}
    \centering
    \includegraphics[width = \linewidth]{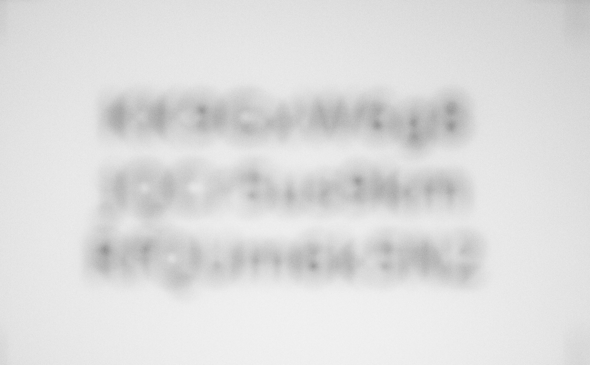}
    %\caption{}
  \end{subfigure}%
  \hspace{.2cm}% Space between image B and C
  \begin{subfigure}{.3\linewidth}
    \centering
    \includegraphics[width = \linewidth]{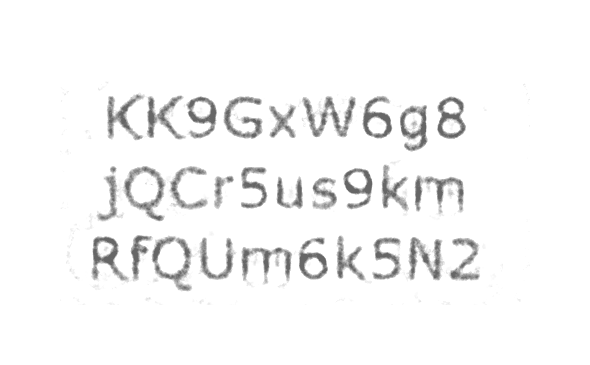}
    %\caption{}
  \end{subfigure} \\
\caption{For each row, from left to right: ground truth, noisy sample and reconstruction using the SVR merit function with $K = 70$ inner iterations. From top to bottom, two images, one per font, are selected for the steps 6 and 8. The corresponding scores are shown in Table \ref{tab:recs_scores}.}
\label{fig:recs_svr_6_8}
\end{figure}

\begin{figure}[!ht] %% Volendo per le ricostruzioni c'è la versione con le bounding box dell'OCR nella cartella figures, ma si vedono poco.
  \centering
  \begin{subfigure}{.3\linewidth}
    \centering
    \includegraphics[width = \linewidth]{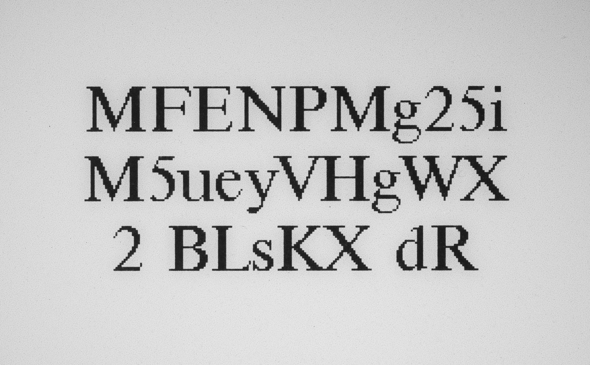}
    %\caption{}
  \end{subfigure}%
  \hspace{.2cm} % Space between image A and B
  \begin{subfigure}{.3\linewidth}
    \centering
    \includegraphics[width = \linewidth]{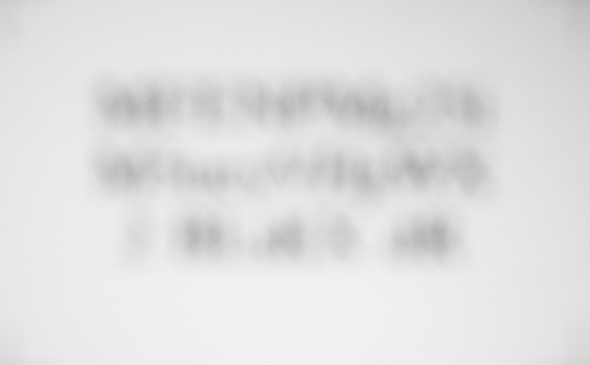}
    %\caption{}
  \end{subfigure}%
  \hspace{.2cm}% Space between image B and C
  \begin{subfigure}{.3\linewidth}
    \centering
    \includegraphics[width = \linewidth]{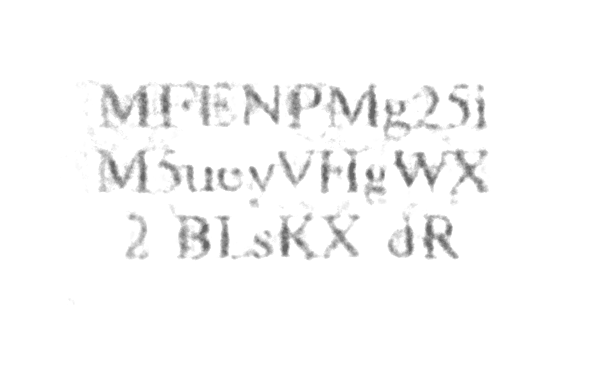}
    %\caption{}
  \end{subfigure} \\
  \vspace{.2cm}
  \begin{subfigure}{.3\linewidth}
    \centering
    \includegraphics[width = \linewidth]{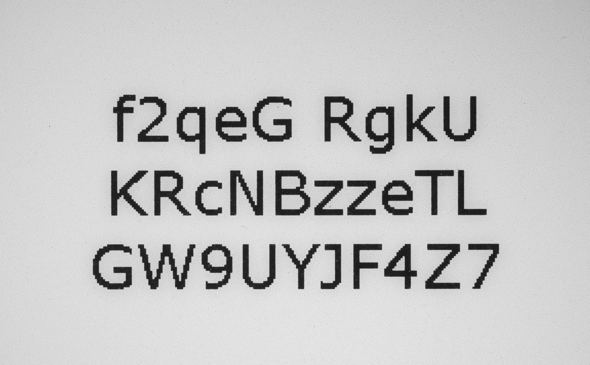}
    %\caption{}
  \end{subfigure}%
  \hspace{.2cm} % Space between image A and B
  \begin{subfigure}{.3\linewidth}
    \centering
    \includegraphics[width = \linewidth]{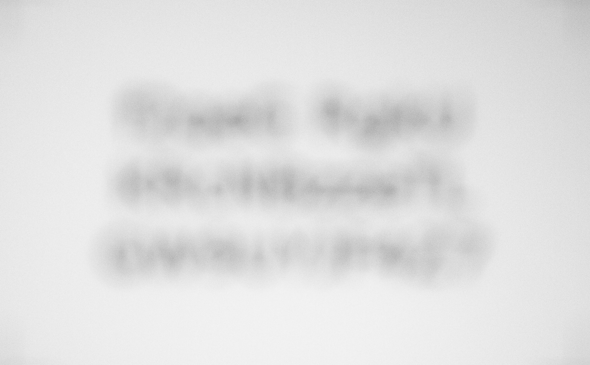}
    %\caption{}
  \end{subfigure}%
  \hspace{.2cm}% Space between image B and C
  \begin{subfigure}{.3\linewidth}
    \centering
    \includegraphics[width = \linewidth]{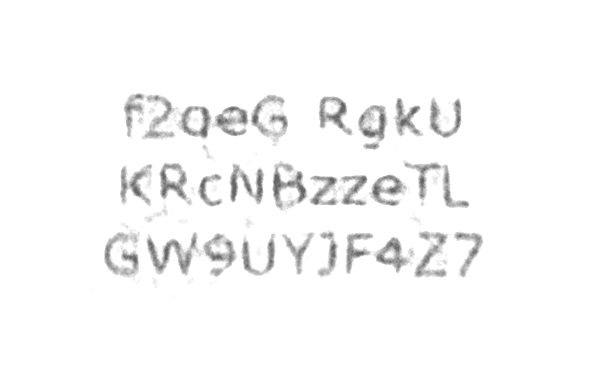}
    %\caption{}
  \end{subfigure} \\
    \vspace{.2cm}
  \begin{subfigure}{.3\linewidth}
    \centering
    \includegraphics[width = \linewidth]{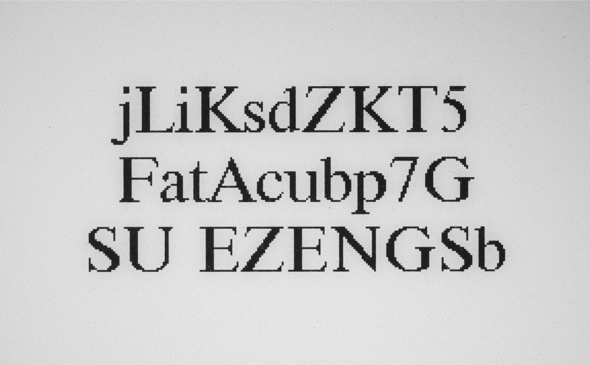}
    %\caption{}
  \end{subfigure}%
  \hspace{.2cm} % Space between image A and B
  \begin{subfigure}{.3\linewidth}
    \centering
    \includegraphics[width = \linewidth]{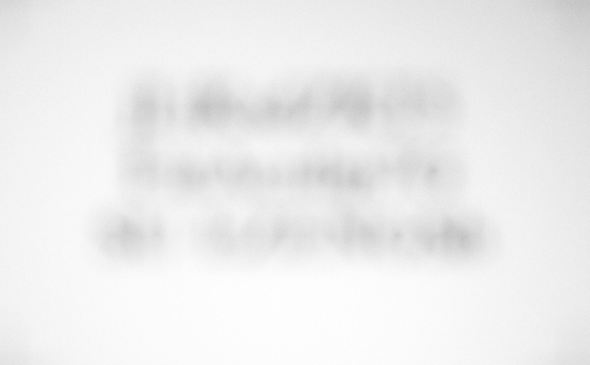}
    %\caption{}
  \end{subfigure}%
  \hspace{.2cm}% Space between image B and C
  \begin{subfigure}{.3\linewidth}
    \centering
    \includegraphics[width = \linewidth]{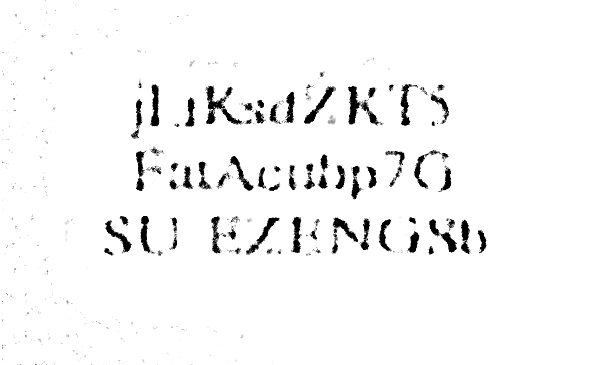}
    %\caption{}
  \end{subfigure} \\
      \vspace{.2cm}
  \begin{subfigure}{.3\linewidth}
    \centering
    \includegraphics[width = \linewidth]{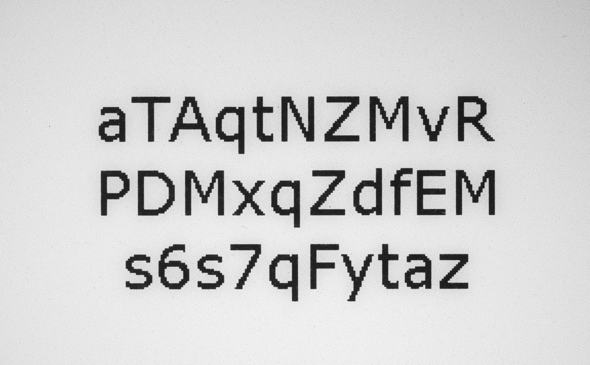}
    %\caption{}
  \end{subfigure}%
  \hspace{.2cm} % Space between image A and B
  \begin{subfigure}{.3\linewidth}
    \centering
    \includegraphics[width = \linewidth]{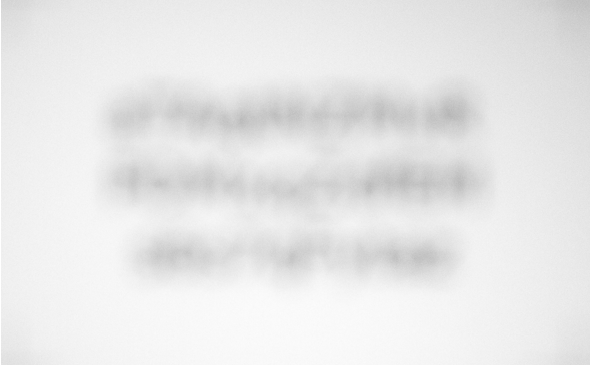}
    %\caption{}
  \end{subfigure}%
  \hspace{.2cm}% Space between image B and C
  \begin{subfigure}{.3\linewidth}
    \centering
    \includegraphics[width = \linewidth]{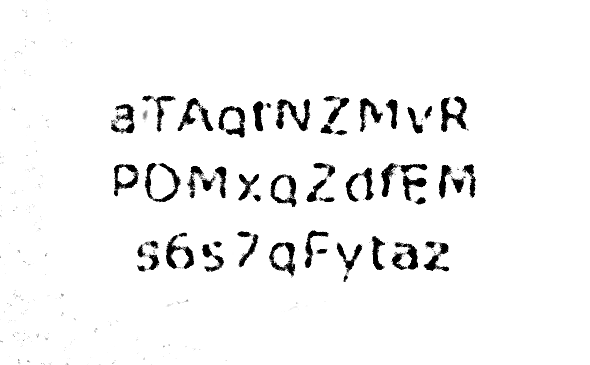}
    %\caption{}
  \end{subfigure} \\
\caption{For each row, from left to right: ground truth, noisy sample and reconstruction using the SVR merit function with $K = 70$ inner iterations. From top to bottom, two images, one per font, are selected for the steps10 and 12. The corresponding scores are shown in Table \ref{tab:recs_scores}.}
\label{fig:recs_svr_10_12}
\end{figure}

In Figure \ref{fig:recs_svr_6_8} and \ref{fig:recs_svr_10_12}  we have reported some examples of reconstructions, after applying the $K = 70$ iterations Algorithm \ref{unrolling} with its parameters learned using the unsupervised approach (the images obtained by minimizing the SSIM-based merit function are visually very similar, thus they have been omitted). In each row the first image is the ground truth, the second is the blurred observation and the third one is our reconstruction with the bounding boxes of the OCR. The SVR and OCR scores are in Table \ref{tab:recs_scores}. Overall, the SVR is a bit conservative with its predictions: it rarely reaches either 100 or 0. However, while this may result is an over/underestimation of the real score, especially for the ground truths and the noisy data, it still manages to discern whether the image is bad or good. It is also worth observing that the predictions attached to the reconstructions are, with different degrees, close to the final average score for the corresponding level (in these examples there is a clear difference for the reconstruction scores because we cherry-picked the images).


\begin{table}
\caption{Comparison between the SVR prediction and the true OCR score for the images of Figure \ref{fig:recs_svr_6_8} and \ref{fig:recs_svr_10_12}.}
\label{tab:recs_scores}
\centering
\begin{tabular}{c | c c | c c | c c}
\toprule
        & \multicolumn{2}{c|}{Ground truth} & \multicolumn{2}{c|}{Noisy sample} & \multicolumn{2}{c}{Reconstruction} \\
        \cline{2-7}
        & OCR & SVR & OCR & SVR & OCR & SVR \\
\midrule
        Step 6 Times & 100 & 81.84 & 0 & 42.17 & 100 & 71.71 \\
        Step 6 Verdana & 100 & 81.91 & 0 & 45.64 & 100 & 74.18 \\
\midrule
        Step 8 Times & 100 & 81.91 & 0 & 34.44 & 90 & 70.1 \\
        Step 8 Verdana & 100 & 81.98 & 0 & 37.68 & 86 & 75.48 \\
\midrule
        Step 10 Times & 100 & 81.92 & 0 & 31.22 & 90 & 66.32 \\ 
        Step 10 Verdana & 100 & 81.96 & 0 & 30.67 & 100 & 68.44 \\
\midrule
        Step 12 Times & 100 & 82.03 & 0 & 24.07 & 76 & 67.42 \\ 
        Step 12 Verdana & 100 & 82.03 & 0 & 29.05 & 70 & 71.53 \\
\bottomrule
\end{tabular}
\end{table}

As a final remark,  we observed that in our method the overfitting phenomenon, as it is commonly meant in machine learning, hardly occurs. This might be due to the fact that the unrolling procedure and the underlying variational model are substantially stiff and they determine a quite resctrictive class of possible reconstructions. On the other side, we believe that this can be also the main limit of the overall approach, since in presence of extremely degraded data and of a coarse approximation of the imaging model, it is not able to provide perfect reconstructions.

\section{Conclusions and future work}

In this paper we collected our research activity carried out on the occasion of the Helsinki Deblur Challenge, in which clean pictures of characters were to be recovered from their out-of-focus photographs in such a way that the characters themselves were identified by an optical character recognition software. The challenge results showed a clear separation between deep learning based approaches and standard variational ones, with the former ones able to reach outstanding performances at the price of an optimization of millions of parameters. One of our purposes was to obtain results comparable to (at least some of) these approaches by keeping an interpretable scheme, in which all the parameters defining the model have a clear meaning and are automatically tuned. To this aim, we proposed a bilevel optimization method, in which the reconstructed image is modelled by unfolding a fixed number of FISTA iterations applied to the minimization of a regularized least-squares functional. In the upper level of the scheme, the parameters defining the functional and the steplengths of the iterations are determined by minimizing a specific merit function, chosen as either a similarity index w.r.t. some ground truth images, or a predictor of the OCR value. A remarkable advantange of this choice is that it completely ignores the ground truth images, thus representing a particularly attractive approach in cases where these information is not available.

The results on the HDC test images show that both the proposed approaches lead to OCR scores comparable to those of some DIP or UNet based algorithms.
We also point out that, although the ingredients and some specific settings we described in the paper have been customized for the HDC images, the proposed scheme can be applied to a general image reconstruction problem, in which different data-fidelity functions or more complex regularizers are needed to provide accurate final images. Our future work will investigate indeed these directions, and we plan both to generalize the energy functional by considering parametric priors as that in \eqref{MRF}, which showed impressive performances in the case of natural images, and to test the scheme for different forward models, as the Radon transform in the case of tomographic images.

% For acknowledgements section, please don't number the section, please begin it with \section*{Acknowledgements}
% \section*{Acknowledgments}

\appendix

\section{Implementation details}

Here we give more details concerned with the computation of the gradient of the merit function w.r.t. to the energy functional parameters $\theta$. Our approach is similar to \cite[Algorithm 1]{Ochs-etal-2015}. To this end, with reference to Algorithm \ref{unrolling}, let us first observe that:
\begin{eqnarray*}
v^{(k)}&=& \Pi((1+\beta_k)u^{(k)} -\beta_k u^{(k-1)}).
\end{eqnarray*}
This in turn implies
\begin{eqnarray*}
\frac{\partial v^{(k)}}{\partial \theta_j}&=& \Pi'(\bar v^{(k)})\left((1+\beta_k)\frac{\partial u^{(k)}}{\partial \theta_j} -\beta_k \frac{\partial u^{(k-1)}}{\partial \theta_j}\right)
\end{eqnarray*}
where, with a little abuse of notation, $\Pi'(v^{(k)})$ formally indicates the diagonal matrix whose entries are obtained by evaluating the derivative of the function \eqref{proj} over the components of the vector $v^{(k)}$.
Then, we have
\begin{eqnarray*}
\frac{\partial u^{(k+1)}}{\partial \theta_j}&=& \Pi'(t^{(k)})\left(\left(I-\alpha_k \nabla_{uu}^2E(v^{(k)},\theta)\right) \frac{\partial v^{(k)}}{\partial \theta_j} -\alpha_k \frac{\partial}{\partial\theta_j}\nabla_u E(v^{(k)},\theta)\right)\\
&=& (1+\beta_k) \Pi'(t^{(k)})\left(I-\alpha_k \nabla_{uu}^2E(v^{(k)},\theta)\right)\Pi'(\bar v^{(k)})\frac{\partial u^{(k)}}{\partial \theta_j}+\\
&&  -\beta_k \Pi'(t^{(k)})\left(I-\alpha_k \nabla_{uu}^2E(v^{(k)},\theta)\right)\Pi'(\bar v^{(k)})\frac{\partial u^{(k-1)}}{\partial \theta_j}+\\
&& -\alpha_k \Pi'(t^{(k)})\frac{\partial}{\partial\theta_j}\nabla_u E(v^{(k)},\theta),
\end{eqnarray*} 
where $t^{(k)}$ is defined as in Algorithm \ref{unrolling}.
The above vector represents the $j$-th column of the Jacobian of the map $u^{(k+1)}=u^{(k+1)}(\theta)$. Using the chain rule for differentiating function composition, we define a matrix-free recursive procedure for computing the gradient of $\ell(u^*(\theta,\alpha))$ with respect to $\theta$. This procedure needs the following initialization:
%\begin{eqnarray*}
%z^{(K+1)} &=& \left.\frac{\partial \ell}{\partial u}\right|_{u=u^*}\\%r^{(K+1)} &=& 0\\
%w_j^{(K+1)} &=& 0
%\end{eqnarray*}
\begin{equation*}
z^{(K+1)} = \frac{\partial \ell}{\partial u^*} \qquad ; \qquad
r^{(K+1)} = 0 \qquad ; \qquad
w_j^{(K+1)} = 0
\end{equation*}
\begin{eqnarray*}
\frac{\partial \ell}{\partial \theta_j} &=&  \left[\frac{\partial u^*}{\partial \theta_j}\right]^T\frac{\partial \ell}{\partial u^*}= \left[\frac{\partial u^{(K+1)}}{\partial \theta_j}\right]^T\frac{\partial \ell}{\partial u^*}\\
																				&=& \left[\frac{\partial u^{(K+1)}}{\partial \theta_j}\right]^Tz^{(K+1)}+ \left[\frac{\partial u^{(K)}}{\partial \theta_j}\right]^Tr^{(K+1)} + w_j^{(K+1)}\\
																				&=& (1+\beta_K)\left[\frac{\partial u^{(K)}}{\partial \theta_j}\right]^T\Pi'(\bar v^{(K)})\left(I-\alpha_K \nabla_{uu}^2E(v^{(K)},\theta)\right)\Pi'(t^{(K)})z^{(K+1)} +\\
																				&& -\beta_K\left[\frac{\partial u^{(K-1)}}{\partial \theta_j}\right]^T\Pi'(\bar v^{(K)})\left(I-\alpha_K \nabla_{uu}^2E(v^{(K)},\theta)\right)\Pi'(t^{(K)})z^{(K+1)}+\\
																				&& -\alpha_K\left[\frac{\partial}{\partial\theta_j}\nabla_u E(v^{(K)},\theta)\right]^T\Pi'(t^{(K)})z^{(K+1)}+\\
																				&& +\left[\frac{\partial u^{(K)}}{\partial \theta_j}\right]^Tr^{(K+1)} + w_j^{(K+1)}
\end{eqnarray*}
Now  we define the following quantities, for $k=1,...,K$
\begin{eqnarray*}
z^{(k)} &=& (1+\beta_k) \Pi'(\bar v^{(k)})\left(I-\alpha_k \nabla_{uu}^2E(v^{(k)},\theta)\right)\Pi'(t^{(k)})z^{(k+1)} + r^{(k+1)}\\
r^{(k)} &=& -\beta_k \Pi'(\bar v^{(k)})\left(I-\alpha_k \nabla_{uu}^2E(v^{(k)},\theta)\right)\Pi'(t^{(k)})z^{(k+1)}\\
w_j^{(k)}&=& -\alpha_k\left[\frac{\partial}{\partial\theta_j}\nabla_u E(v^{(k)},\theta)\right]^T\Pi'(t^{(k)})z^{(k+1)} + w_j^{(k+1)}.
\end{eqnarray*}
Then,
\begin{eqnarray*}
\frac{\partial \ell}{\partial \theta_j} &=& \left[\frac{\partial u^{(K)}}{\partial \theta_j}\right]^Tz^{(K)}+ \left[\frac{\partial u^{(K-1)}}{\partial \theta_j}\right]^Tr^{(K)} + w_j^{(K)}.
\end{eqnarray*}
We are now ready to completely state the recursion procedure for computing the gradient of the merit function, which is detailed in Algorithm \ref{loss_gradient}.
\begin{algorithm}
\caption{Loss function gradient computation by backward differentiation.}\label{loss_gradient}
\textsc{Input}: $u^*$ from Algorithm \ref{unrolling}.
%\begin{equation}\nonumber
$
\begin{array}{ll}
z^{(K+1)} = \dfrac{\partial \ell}{\partial u^*}\\
r^{(K+1)} = 0\\
w_j^{(K+1)} = 0\\
\mbox{\textsc{For} }  k = K,K-1,...,1\\
           \left\lfloor\begin{array}{lcl}
						   \tilde z^{(k)} &=& \Pi'(t^{(k)})z^{(k+1)}\\[0.2cm]
							 w_j^{(k)}     &=& -\alpha_k\left[\frac{\partial}{\partial\theta_j}\nabla_u E(v^{(k)},\theta)\right]^T\Pi'(t^{(k)})z^{(k+1)} + w_j^{(k+1)}\\[0.2cm]
							 q^{(k)}     &=& \Pi'(\bar v^{(k)})\left(I-\alpha \nabla_{uu}^2E(v^{(k)},\theta)\right)\tilde z^{(k)}\\[0.2cm]
							z^{(k)}    &=& (1+\beta_k) q^{(k)} + r^{(k+1)}\\[0.2cm]
							r^{(k)}     &=& -\beta_k q^{(k)}
					    \end{array}\right.\\
\displaystyle\frac{\partial \ell}{\partial \theta_j} = w_j^{(1)} - \alpha_0\left[\frac{\partial}{\partial\theta_j}\nabla_u E(v^{(0)},\theta)\right]^T\Pi'(t^{(0)})z^{(1)}					
\end{array}
%\end{equation}
$\\
\textsc{Output}: $\displaystyle\frac{\partial \ell}{\partial \theta_j}$
\end{algorithm}

\section{Application on natural images}

The HDC rules declared in the challenge web page \href{https://fips.fi/HDCrules.php}{https://fips.fi/HDCrules.php} required that the competing algorithms are actual deconvolution methods: approaches whose output is always text, regardless of the input data, were excluded by the challenge. To this end, before the testing phase, each algorithm was applied to some technical targets and natural images as input data. This stage, called sanity check, was passed if the algorithm output was even a slightly deblurred version of the input image. \\
In this section we show the results of our method applied to some of the natural images belonging to the sanity test set (see Figure \ref{fig:safety}) . These results are obtained by applying Algorithm \ref{unrolling} to the blurred noisy input image, with the same parameters $\theta,\alpha$ learned as described in sections \ref{sec_loss}-\ref{sec_tests}. Notice that the training set used for the parameters tuning was composed only by text images. Moreover, the energy functional in our approach, mainly in its regularization part, is specifically tailored for restoring sparse, binary, piecewise constant images. This explains why the output of Algorithm \ref{unrolling} on natural images is not much satisfactory, while a more acceptable result is obtained on the QR code and on the dandelion image, which are quite similar to text images. \\
Indeed, even if the main idea of bilevel/unrolling technique is very general and can be applied to a variety of image restoration problems, the design of the energy functional must be adapted to the specific features of the kind of images it is designed for. As for the upper level problem, the SSIM loss function can be used as it is for training models suited for natural images, as proposed also in \cite{Bertocchi-etal-2019}. On the other side, also the SVR loss function can still be generalized to different contexts by replacing the OCR score, which is specific for the HDC application, with other quality measures, as for example the PSNR \cite{Li-etal-2017}. This possibility will be subject of future work. 
\def\scalfig{0.045}
\begin{figure}
    \centering
    \begin{tabular}{cccc}
    \includegraphics[scale=\scalfig]{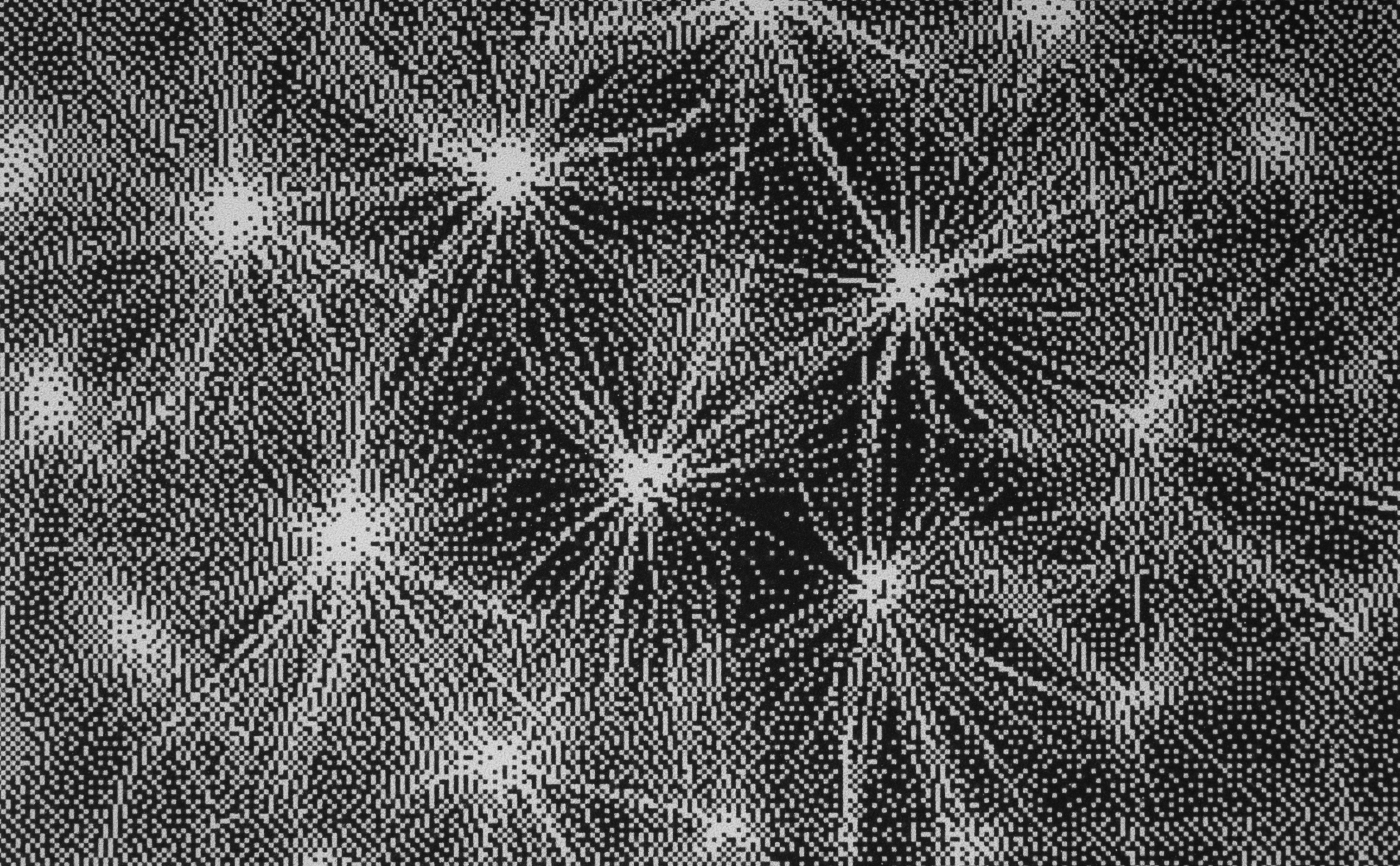}&
    \includegraphics[scale=\scalfig]{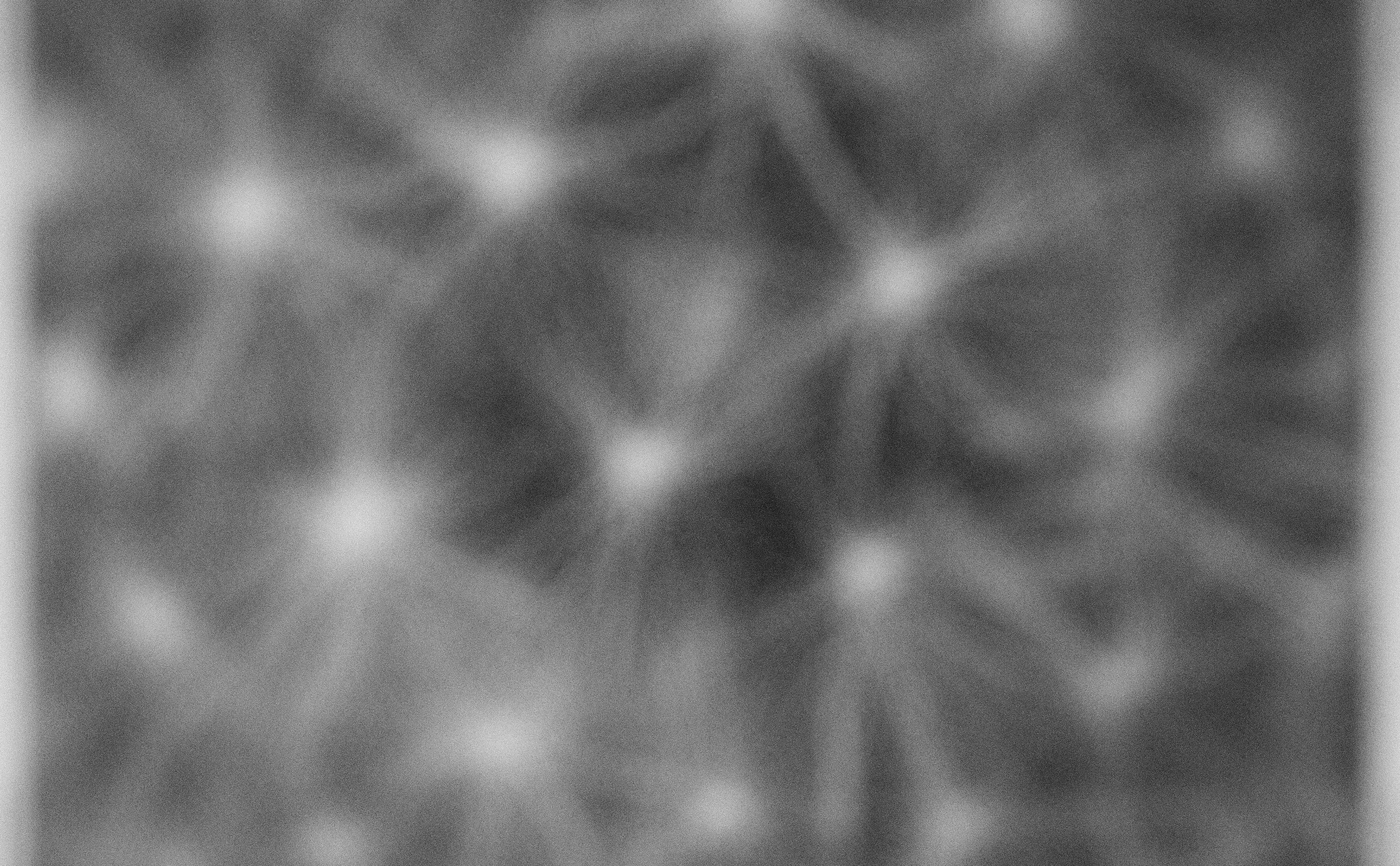}
    &\includegraphics[scale=\scalfig]{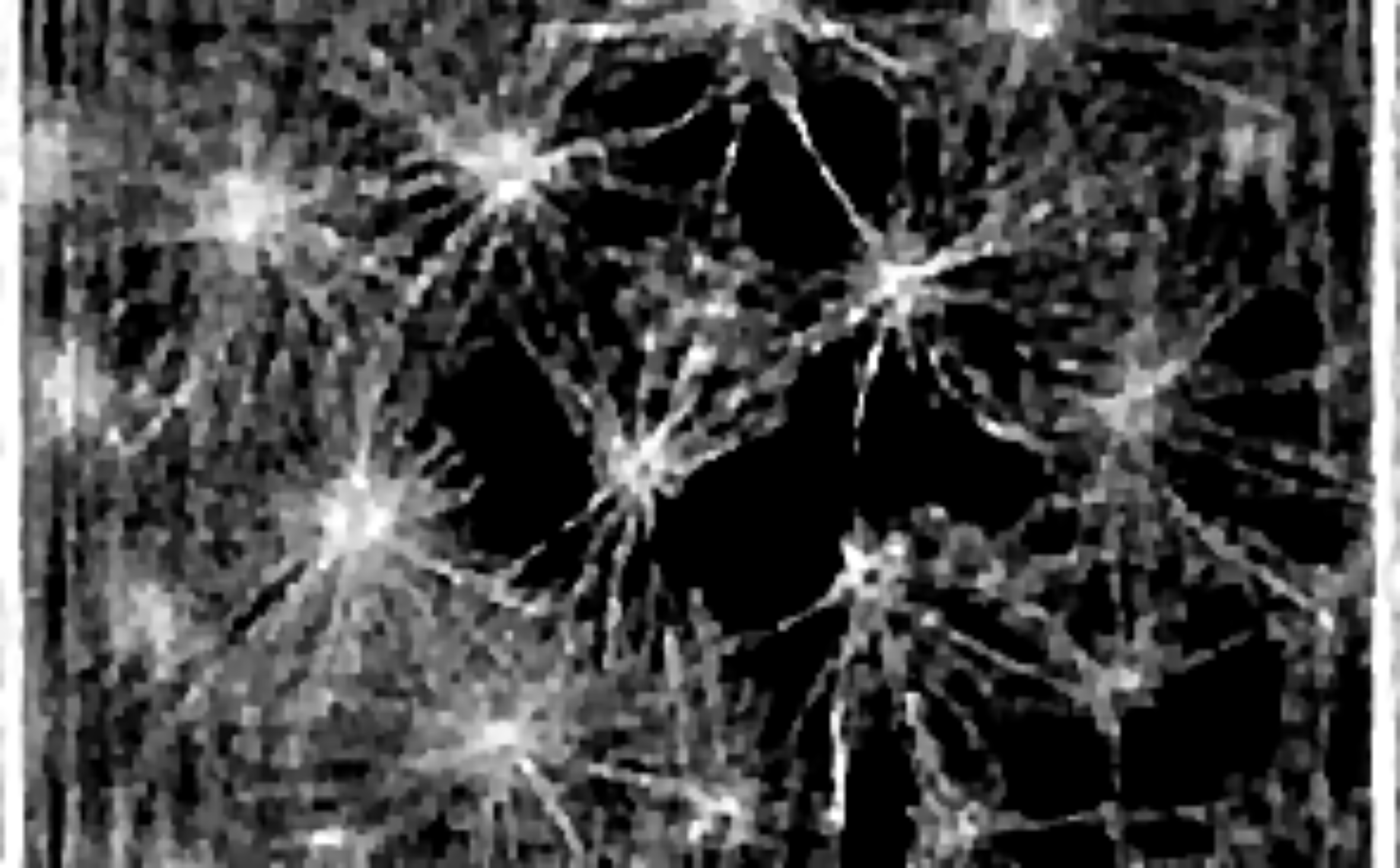}\\
    \includegraphics[scale=\scalfig]{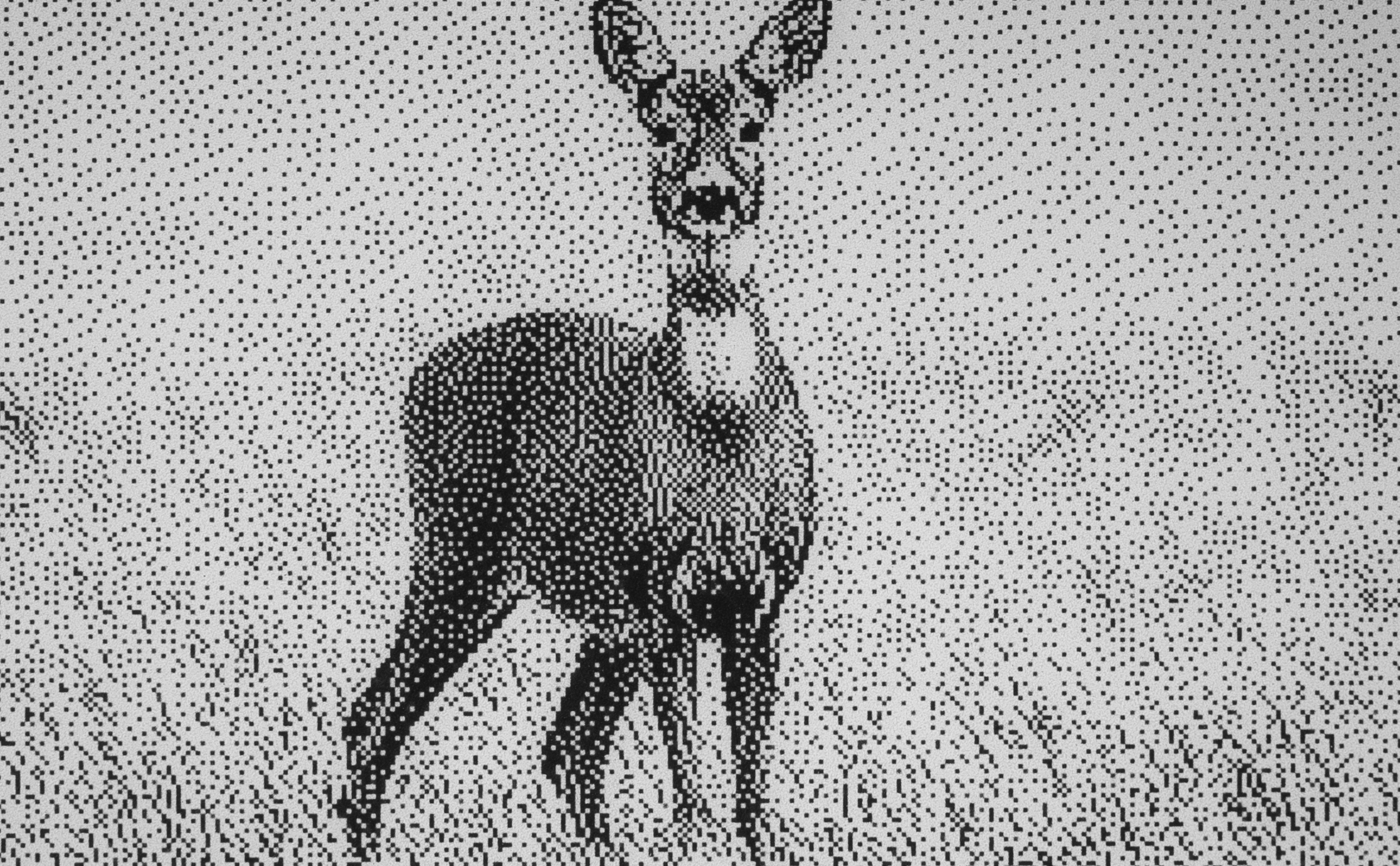}&
    \includegraphics[scale=\scalfig]{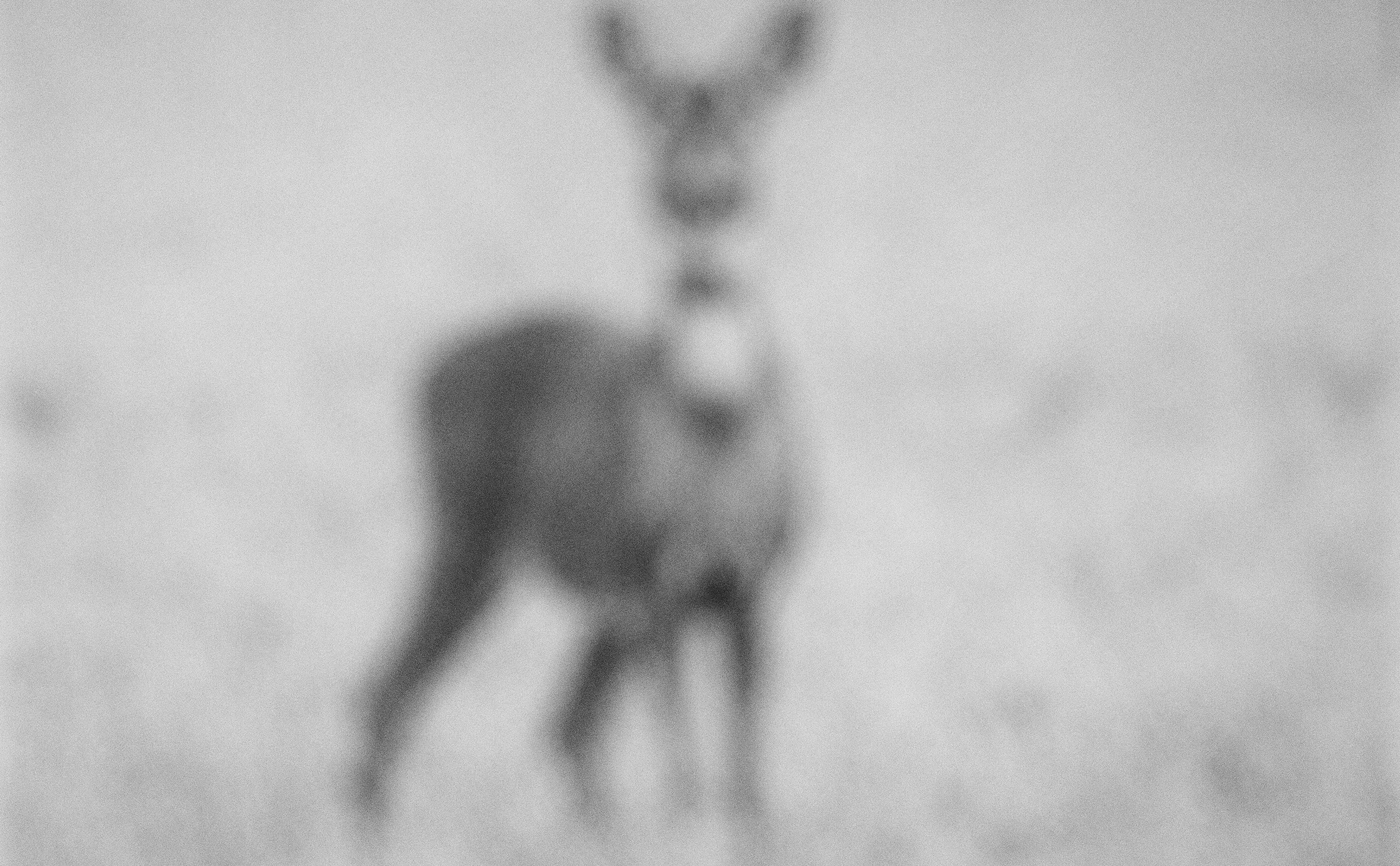}
    &\includegraphics[scale=\scalfig]{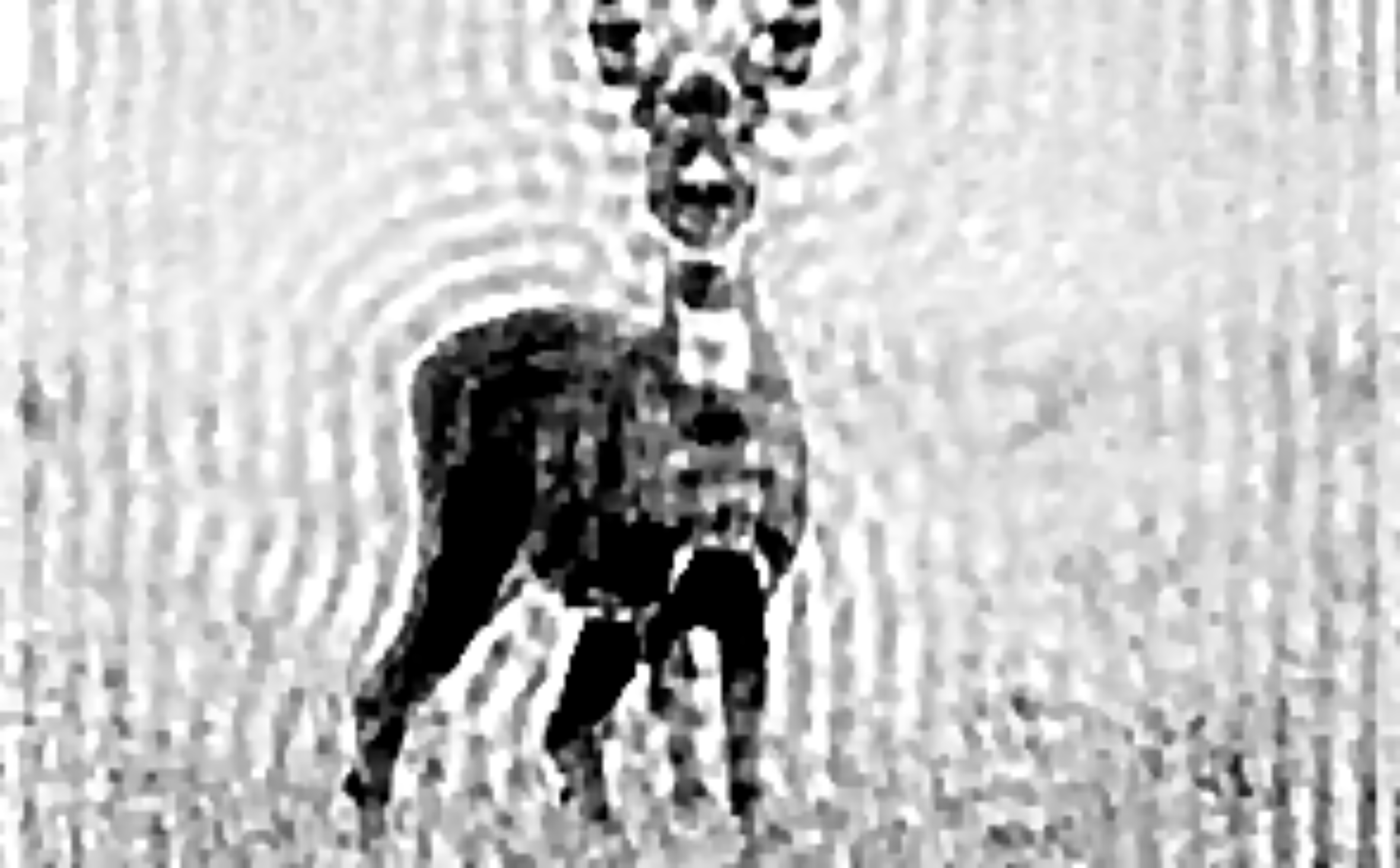}\\
    \includegraphics[scale=\scalfig]{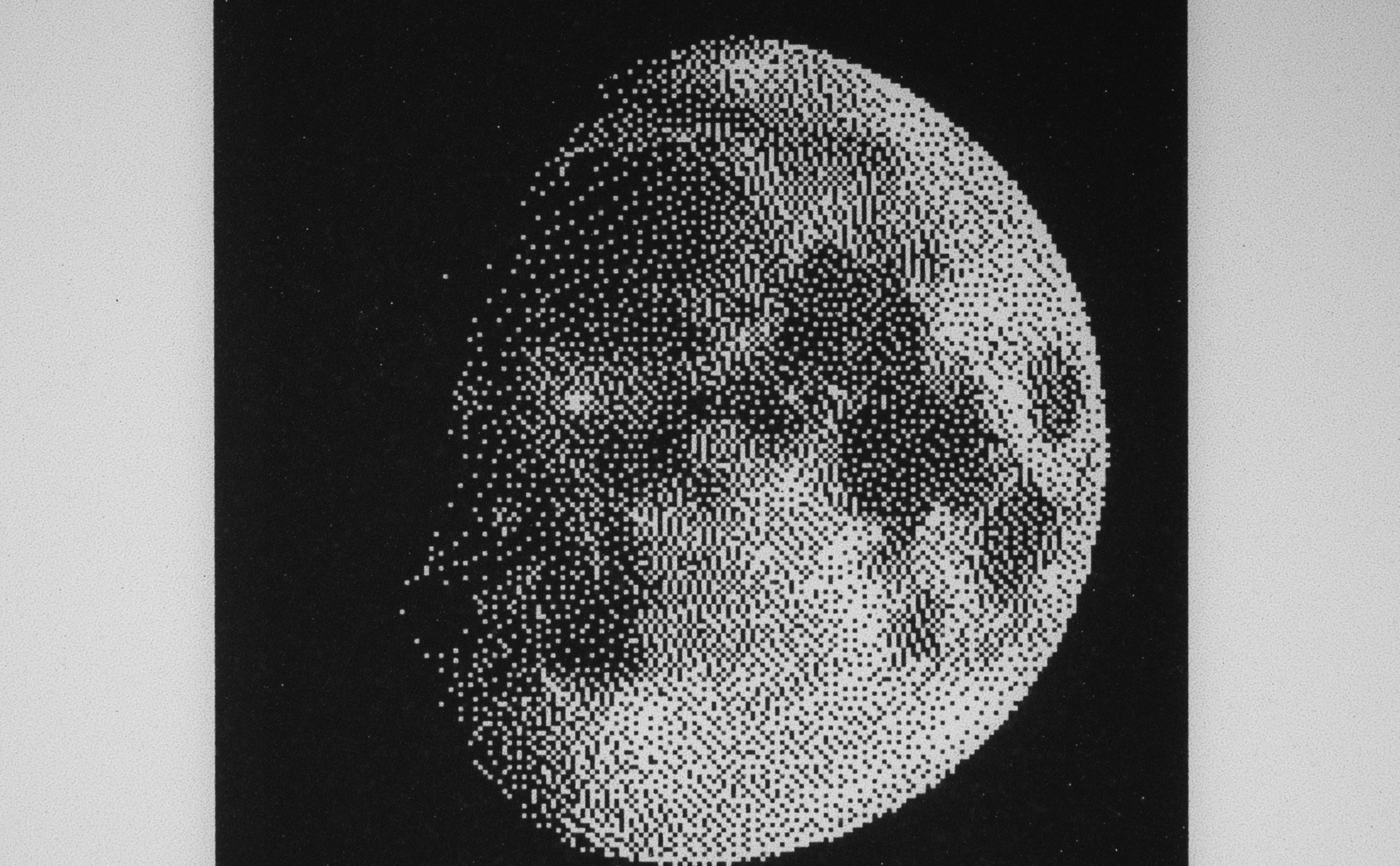}&
    \includegraphics[scale=\scalfig]{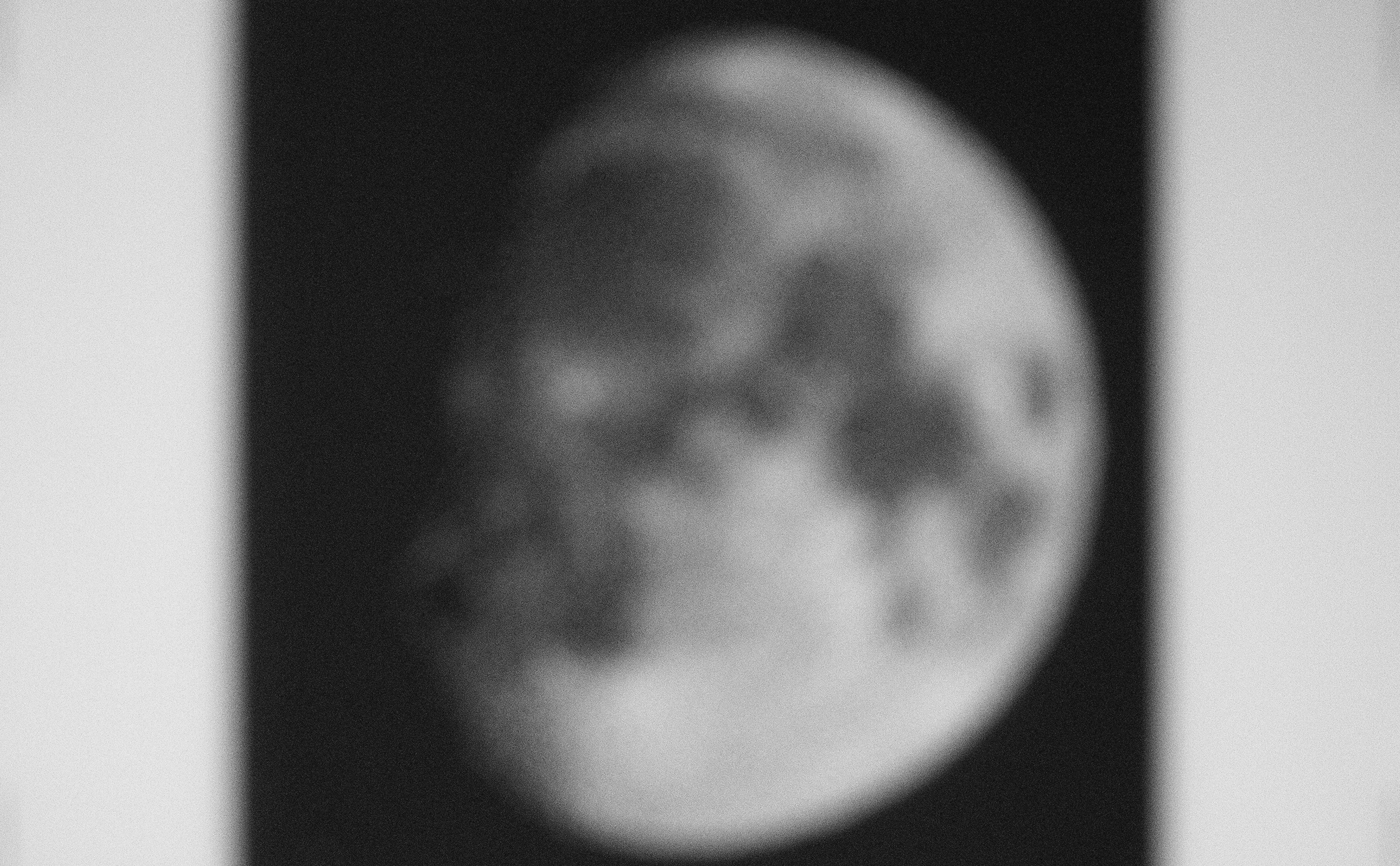}
    &\includegraphics[scale=\scalfig]{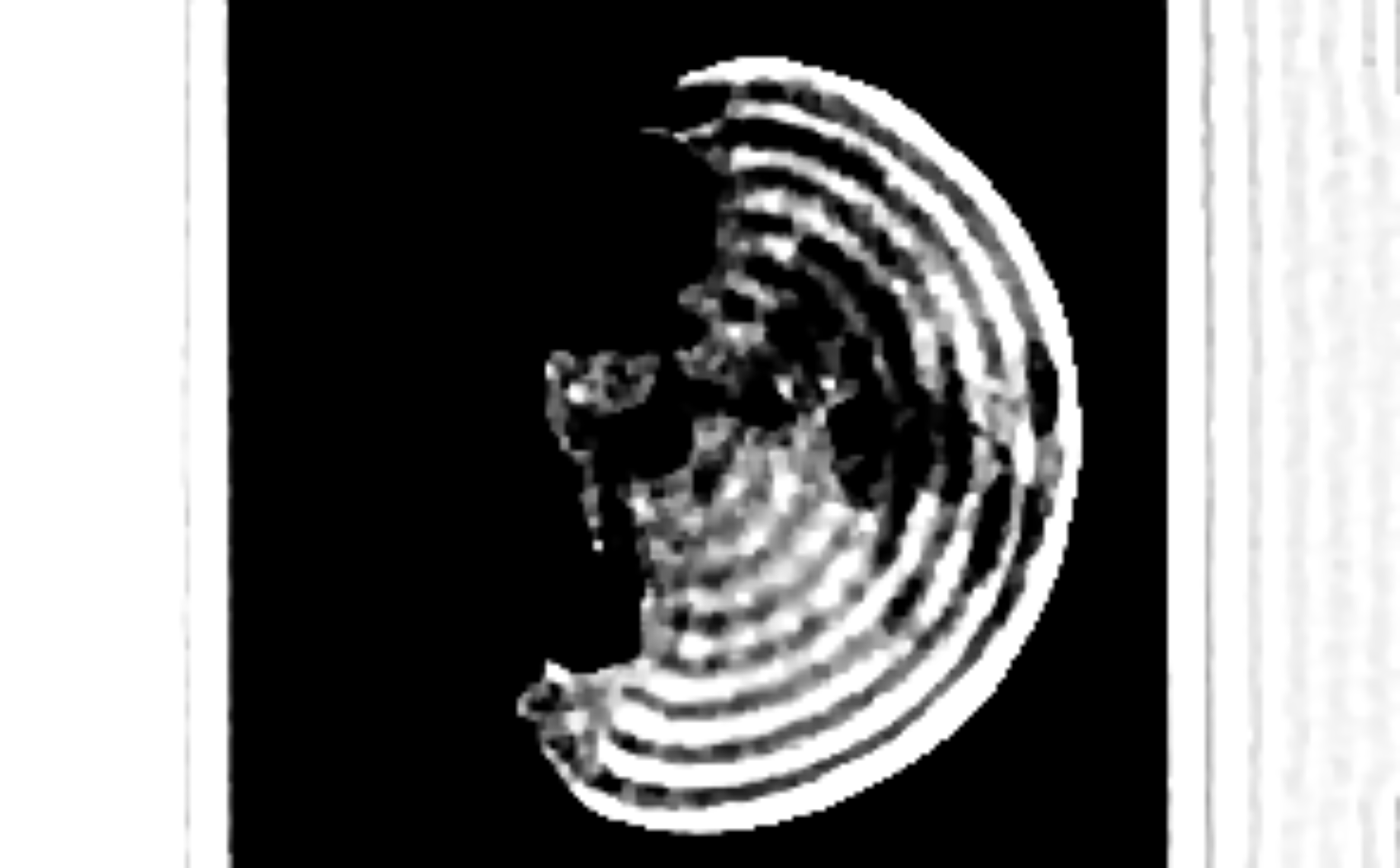}\\
    \includegraphics[scale=\scalfig]{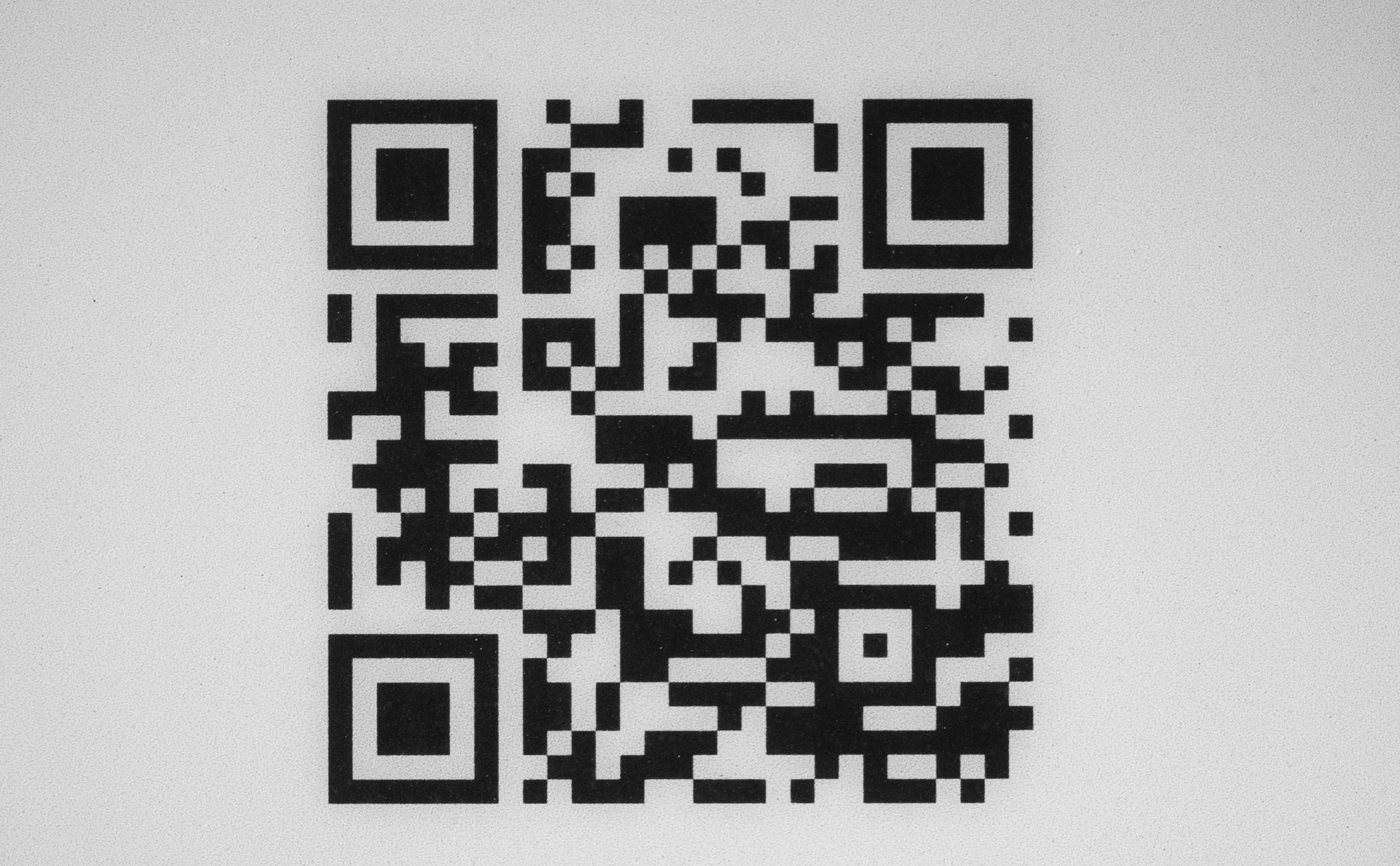}&
    \includegraphics[scale=\scalfig]{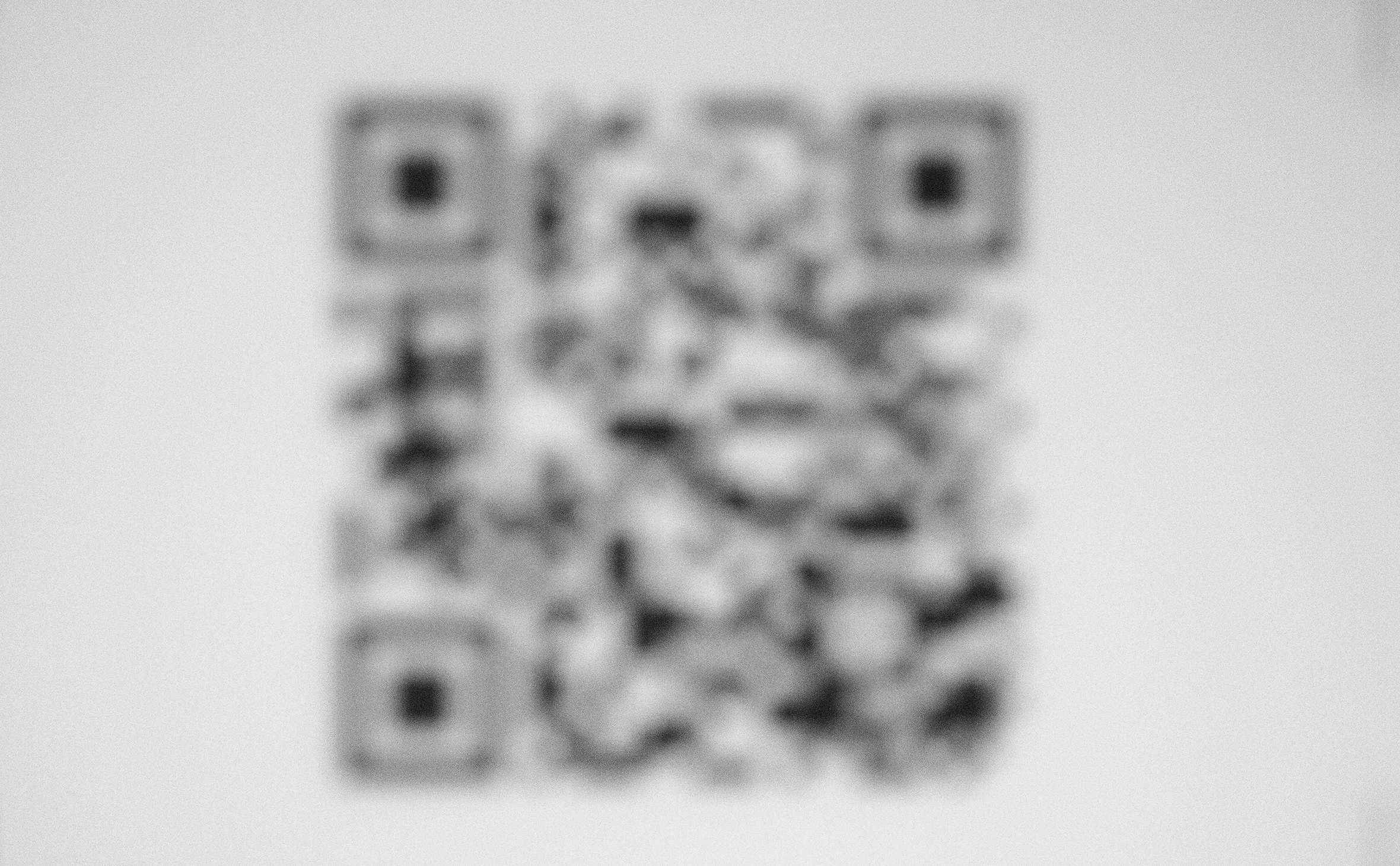}
    &\includegraphics[scale=\scalfig]{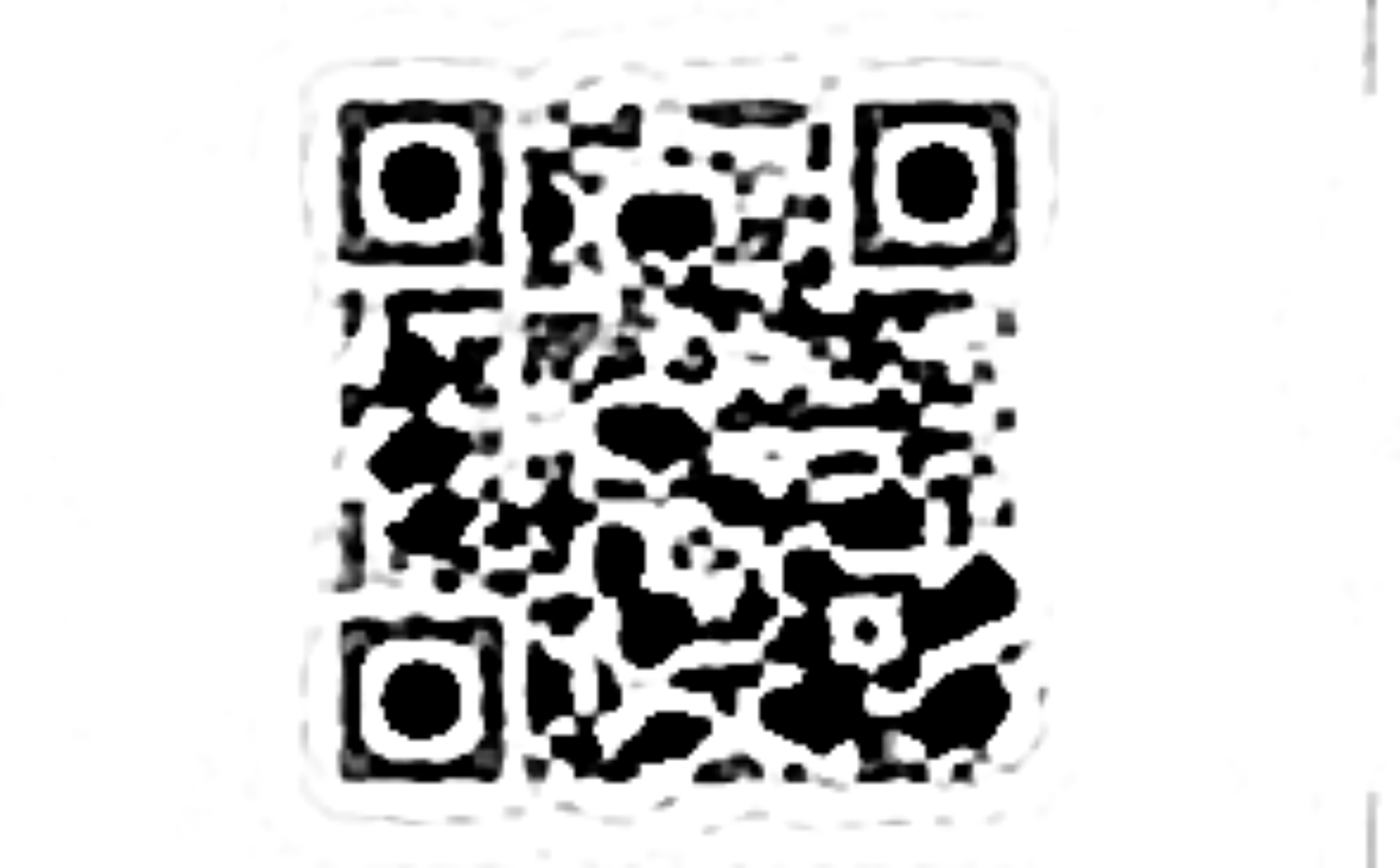}
    \end{tabular}
    \caption{Results of Algorithm \ref{unrolling} with the SSIM loss function and 60 unrolled iterations applied to some of the test images employed for the sanity check (step 6).}
    \label{fig:safety}
\end{figure}

%\medskip
% The data information below will be filled by AIMS editorial staff
%Received xxxx 20xx; revised xxxx 20xx.
%\medskip

\end{document}